\documentclass[a4paper,11pt]{article}
\pdfoutput=1 

\usepackage{jcappub} 

\usepackage[T1]{fontenc} 

\usepackage{lscape}

\usepackage{lineno}

\usepackage{verbatim}

\usepackage[3D]{movie15}
\usepackage{hyperref}
\usepackage[UKenglish]{babel}

\newcommand{\dru}{($\rm{kg} \cdot day \cdot keV)^{-1}$}
\newcommand{\tyu}{($\rm{t} \cdot y)^{-1}$}
\newcommand{\mean}[1]{\langle #1\rangle}

\newcommand\Tstrut{\rule[0pt]{0pt}{2.6ex}}         
\newcommand\Bstrut{\rule[-0.9ex]{0pt}{0pt}}   

\title{Physics reach of the XENON1T dark matter experiment.}

\author{The XENON collaboration: \\}
\author[a]{E.~Aprile,}
\author[b]{J.~Aalbers,}
\author[c,d]{F.~Agostini,}
\author[e]{M.~Alfonsi,}
\author[f]{F.~D.~Amaro,}
\author[a]{M.~Anthony,}
\author[g]{L.~Arazi,}
\author[h]{F.~Arneodo,}
\author[f]{C.~Balan,}
% ...

\affiliation[a]{Physics Department, Columbia University, New York, NY, USA}
\affiliation[b]{Nikhef and the University of Amsterdam, Science Park, Amsterdam, Netherlands}
\affiliation[c]{INFN-Laboratori Nazionali del Gran Sasso and Gran Sasso Science Institute, L'Aquila, Italy}
\affiliation[d]{Department of Physics and Astrophysics, University of Bologna and INFN-Bologna, Bologna, Italy}
\affiliation[e]{Institut f\"ur Physik \& Exzellenzcluster PRISMA, Johannes Gutenberg-Universit\"at Mainz, Mainz, Germany}
\affiliation[f]{Department of Physics, University of Coimbra, Coimbra, Portugal}
\affiliation[g]{Department of Particle Physics and Astrophysics, Weizmann Institute of Science, Rehovot, Israel}
\affiliation[h]{New York University Abu Dhabi, Abu Dhabi, United Arab Emirates}
\affiliation[i]{Physik-Institut, University of Zurich, Zurich, Switzerland}
\affiliation[j]{Department of Physics, Stockholm University, AlbaNova, SE-106 91 Stockholm, Sweden}
\affiliation[k]{Department of Physics, Applied Physics and Astronomy, Rensselaer Polytechnic Institute, Troy, NY, USA}
\affiliation[l]{Max-Planck-Institut f\"ur Kernphysik, Heidelberg, Germany}
\affiliation[m]{Institut f\"ur Kernphysik, Westf\"alische Wilhelms-Universit\"at M\"unster, M\"unster, Germany}
\affiliation[n]{Albert Einstein Center for Fundamental Physics, University of Bern, Bern, Switzerland}
\affiliation[o]{Department of Physics and Astronomy, Purdue University, West Lafayette, IN, USA}
\affiliation[p]{Subatech, Ecole des Mines de Nantes, CNRS/In2p3, Universit\'e de Nantes, Nantes, France}
\affiliation[q]{INFN-Torino and Osservatorio Astrofisico di Torino, Torino, Italy}
\affiliation[r]{Physics \& Astronomy Department, University of California, Los Angeles, CA, USA}
\affiliation[s]{Department of Physics and Astronomy, Rice University, Houston, TX, USA}

\emailAdd{cyril.grignon@uni-mainz.de}
\emailAdd{marco.selvi@bo.infn.it}

\abstract{
The XENON1T experiment is currently in the commissioning phase at the Laboratori Nazionali del Gran Sasso, Italy.
In this article we study the experiment's expected sensitivity to the spin-independent WIMP-nucleon interaction cross section, based on Monte Carlo predictions of the electronic and nuclear recoil backgrounds.

The total electronic recoil background in $1$~tonne fiducial volume and ($1$, $12$)~keV electronic recoil equivalent energy region, before applying any selection to discriminate between electronic and nuclear recoils,  is $(1.80 \pm 0.15) \cdot 10^{-4}$~\dru, mainly due to the decay of $^{222}\rm{Rn}$ daughters inside the xenon target. The nuclear recoil background in the corresponding nuclear recoil equivalent energy region ($4$, $50$)~keV, is composed of $(0.6 \pm 0.1)$~\tyu $\,$ from radiogenic neutrons, $(1.8 \pm 0.3) \cdot 10^{-2}$~\tyu $\,$ from coherent scattering of neutrinos, and less than $0.01$~\tyu $\,$ from muon-induced neutrons.

The sensitivity of XENON1T is calculated with the Profile Likelihood Ratio method, after converting the deposited energy of electronic and nuclear recoils into the scintillation and ionization signals seen in the detector. 
We take into account the systematic uncertainties on the photon and electron emission model, and on the estimation of the backgrounds, treated as nuisance parameters. The main contribution comes from the relative scintillation efficiency $\mathcal{L}_\mathrm{eff}$, which affects both the signal from WIMPs and the nuclear recoil backgrounds. After a $2$~y measurement in $1$~t fiducial volume, the sensitivity reaches a minimum cross section of $1.6 \cdot 10^{-47}$~cm$^2$ at m$_\chi$=$50$~GeV/$c^2$. 

}


\begin{document}
\maketitle
\flushbottom


\section{Introduction}
\label{S:Intro}

Astronomical and cosmological observations indicate
that a large fraction of the energy content of the Universe
is composed of cold dark matter \cite{PDG2014, Blum, Davis, Clowe}.
Recently, increasingly detailed studies of the cosmic microwave background anisotropies have inferred the abundance
of dark matter with remarkable precision at $(26.0 \pm 0.5)\%$ \cite{Bennett, Ade:2015xua}. 
One of the most favored particle candidate, under
the generic name of Weakly Interacting Massive Particles
(WIMPs), arises naturally in many theories beyond the
Standard Model of particle physics, such as supersymmetry, universal extra dimensions, or little Higgs models \cite{Steigman, Jungman, BertoneBook}. Although other candidates exist, like axions \cite{Visinelli:2009zm}, superheavy particles \cite{Chung:1998zb} and sterile neutrinos \cite{Boyarsky:2009ix}, in this study we will focus on the detection of WIMPs.



Among the various experimental strategies to directly detect WIMP interactions,
detectors using liquid xenon (LXe) as target have demonstrated the highest sensitivities over the past
years, for WIMP mass $m_{\chi} > 6$ GeV/$c^2$ \cite{Undagoitia:2015gya}. 
In 2012 and 2013 the XENON100 experiment published the world's best upper limits on
the spin-independent \cite{xe100-run10} and spin-dependent \cite{xe100-sd} coupling of 
WIMPs to nucleons and neutrons, respectively. 
At the end of 2013, the spin-independent result was confirmed and improved by
the LUX experiment  \cite{LUX-results}, also using LXe.
To significantly increase the sensitivity with respect to the current scenario, the XENON
collaboration is focusing on the XENON1T experiment \cite{XENON1T}. The detector construction
in Hall B of the Laboratori Nazionali del Gran Sasso (LNGS) has started in summer 
2013. At the end of 2015 the detector commissioning begun and the first science
run is expected in the first months of 2016.
With a target mass $32$ times larger than XENON100 and a reduced background
rate, the sensitivity to the spin-independent WIMP-nucleon interaction cross section is expected to improve by two orders of magnitude with respect to the XENON100 limits.

A robust estimation of the background rate of the XENON1T experiment is a key ingredient for the sensitivity estimation.
We can divide the background sources in two main classes: electronic recoils (ER) off the atomic electrons and nuclear recoils (NR) off the Xe nuclei. The ER background is from radioactivity in the detector materials, sources intrinsic to the LXe (beta decay of $^{85}\rm{Kr}$, of $^{222}\rm{Rn}$ and its daughters, and $^{136}\rm{Xe}$ double-beta decay) and from solar neutrinos scattering off electrons.
The NR background comes from neutrons originated in spontaneous fission, ($\alpha$, n) reactions and muon-induced interactions (spallations, photo-nuclear and hadronic interactions). Neutrinos, in particular those from the $^{8}\rm{B}$ channel in the Sun, also contribute to the NR background through coherent neutrino-nucleus scattering, an interaction predicted in the Standard Model though not yet observed \cite{Freedman:1973yd}. 
The WIMP signal is also expected to be of the NR type, with a single scatter uniformly distributed in the target volume.

Up to the current generation of direct-search dark matter experiments, ERs constitute the main source of background. It has been very well characterized through Monte Carlo (MC) simulations, with measurements in good agreement with the predictions, see for instance the studies in XENON100 \cite{xe100-erbkg}, EDELWEISS \cite{Edelweiss} and LUX \cite{LUX-background}. The NR background has been predicted \cite{xe100-nrbkg, Edelweiss, LUX-background}, but no direct measurement has been possible so far, given the small number of expected NR events in the exposure of the current experiments.

The goal of the XENON1T experiment requires an ultra-low background (order of a few $10^{-4}$~\dru $\,$ for the total ER background) in the central detector region. Hence we performed a careful screening and selection campaign
for all the detector construction materials, especially those in close proximity to the
xenon target, and we developed powerful purification techniques to remove the intrinsic contaminants from the xenon. 
The external gammas and neutrons from the muons and laboratory environment are reduced to negligible levels by operating the experiment deep underground at LNGS and by placing the detector inside a shield made of at least $4$~m of water, contained in a stainless steel tank.
The water tank is instrumented with light sensors to be operated as a Cherenkov muon veto \cite{XENON1T-MuonVeto}.


The aim of this paper is to describe the various sources of background in XENON1T, 
quantify their contribution together with the associated fluctuations when converted into the signals observed in the detector, 
and calculate the sensitivity of the experiment in the search for WIMP interactions.
The work is organized as follows: in section \ref{S:Geom} we describe the details of the MC simulation of the detector and the contaminations of the materials used to build the experiment. The ER and NR background predictions are described in sections \ref{S:Em} and \ref{S:Nr}, respectively. In section \ref{S:LCE} we describe the simulation of the light collection efficiency of the detector. The conversion from the energy deposited in LXe to the light and charge signals in the detector is described in section  \ref{S:Conv}. The physics reach of the XENON1T experiment and of its upgraded version XENONnT are discussed in section \ref{S:Sens}. Finally, summary and conclusions are presented in section \ref{S:Summ}.

\section{The XENON1T experiment and its simulation in GEANT4}
\label{S:Geom}

\begin{figure}[t!]
\centering\includegraphics[width=0.45\linewidth]{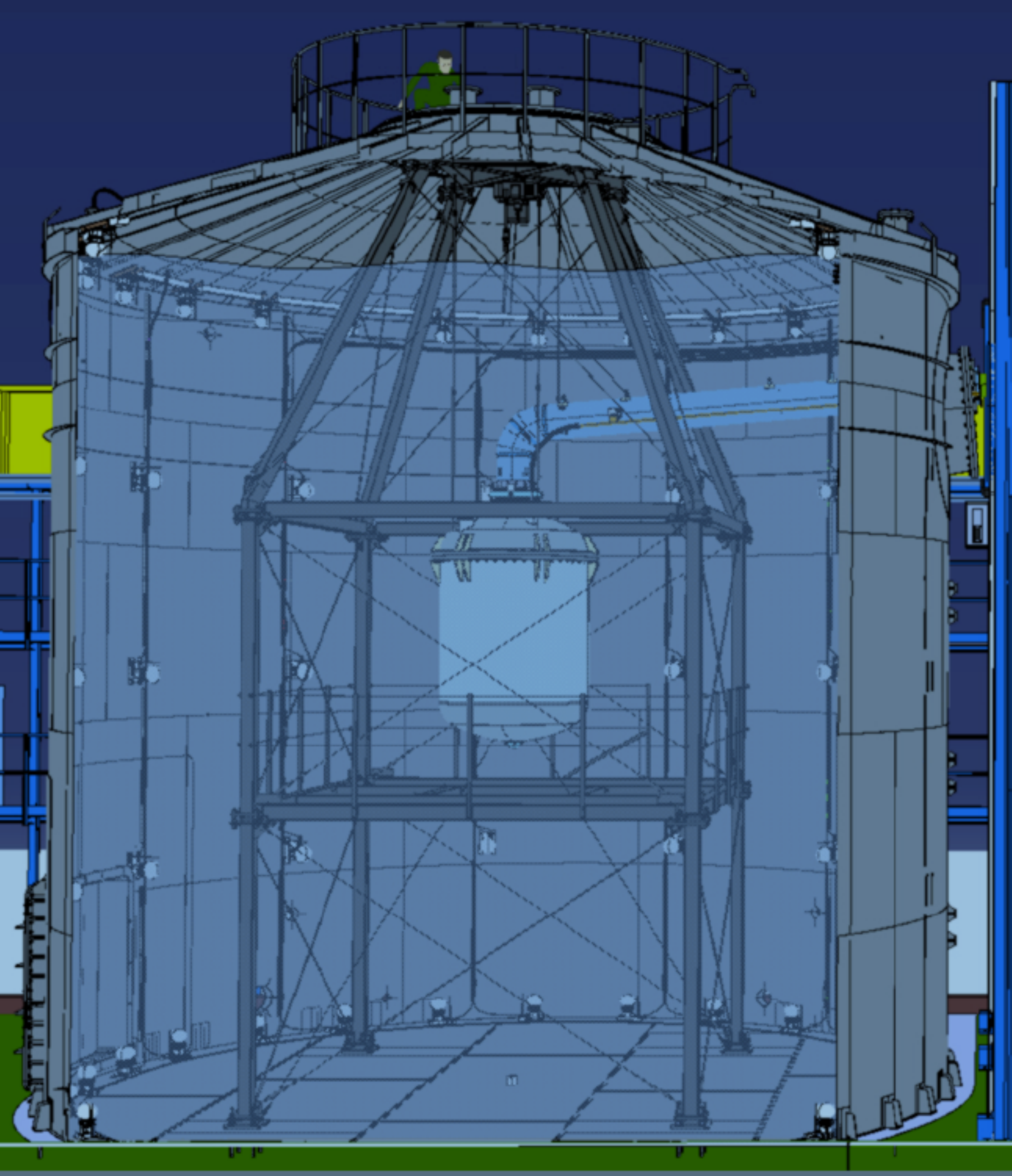}
\centering\includegraphics[width=0.54\linewidth]{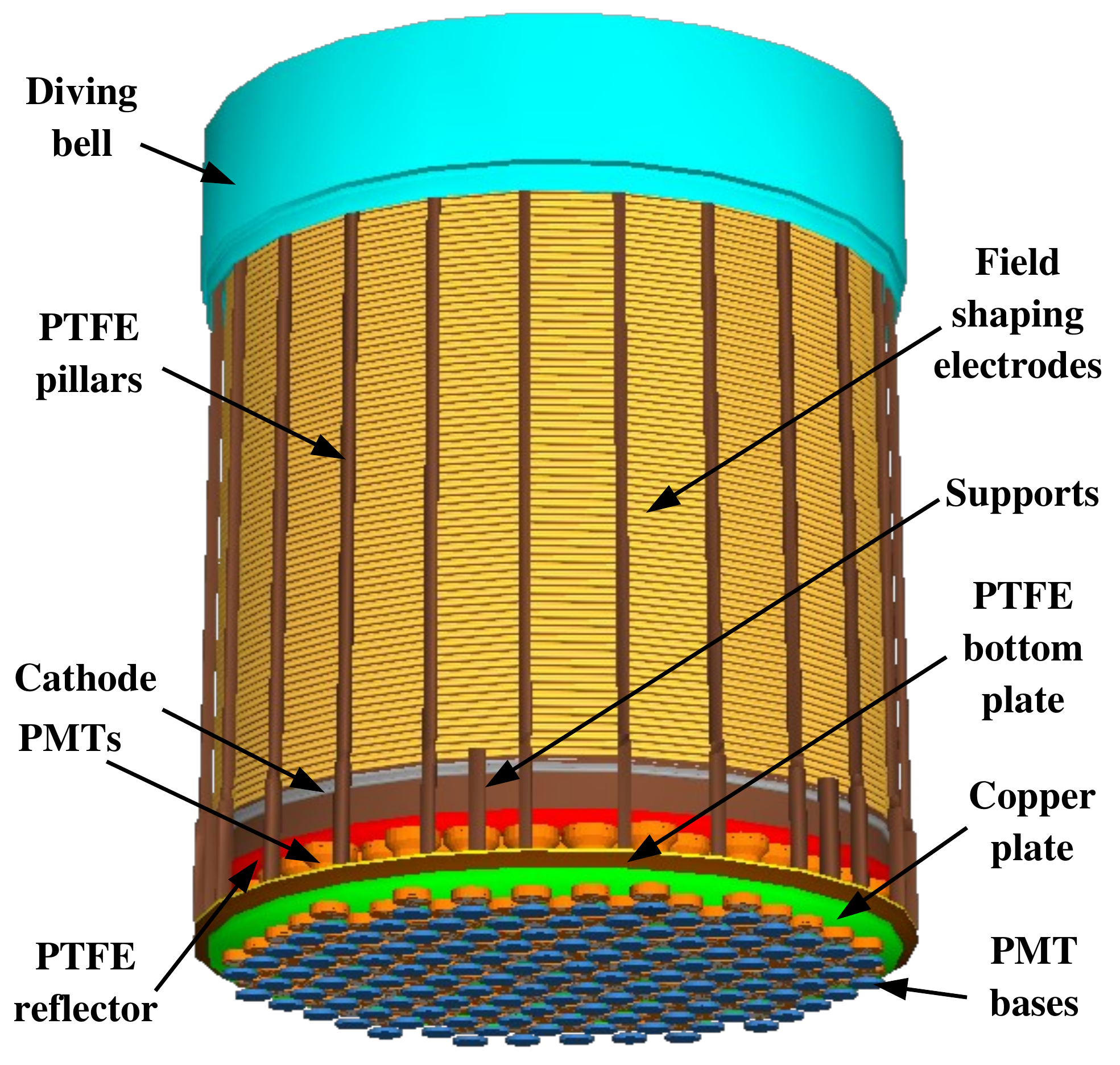}
\caption{(Left) Artistic view showing the double-wall stainless steel cryostat positioned in the center of the water tank. Also visible are the PMTs of the muon veto system, the support structure and the main pipe. (Right) Rendering of the external part of the TPC, as modeled in the GEANT4 simulation. From top to bottom, the diving bell (cyan), PTFE pillars and support parts (brown), field shaping electrodes (yellow), PTFE reflector among the PMTs (red), PMTs in the bottom array (orange), copper plate (green), PMT bases (blue). }
\label{fi:detector}
\end{figure}

The XENON1T detector is a dual phase time projection chamber (TPC) \cite{Bolozdynya:1999} filled with xenon. A particle interaction in the LXe target produces both excited and ionized atoms. De-excitation of excited molecular states leads to a prompt scintillation signal (S1), which is recorded by photomultiplier tubes (PMTs) placed below the target in the LXe and above it in the gaseous phase (GXe). Applying an electric drift field (the design goal is to reach $1$~kV/cm), a large fraction of the ionization electrons is moved away from the interaction site to the top of the TPC. Here the electrons are extracted by a strong electric field (of the order of $10$~kV/cm) from the liquid into the gaseous phase creating a secondary scintillation signal (S2) by collisions with xenon atoms, which is proportional to the number of extracted ionization electrons \cite{Lansiart:1976}. The S2 signal is detected by the same PMTs, and it is delayed with respect to S1 by the time required to drift the electron cloud from the interaction site to the liquid/gas interface. 
With the information recorded in the TPC, a 3-dimensional vertex reconstruction is possible using the drift time for the vertical coordinate and the hit pattern of the S2 signal on the top PMT array for the (x, y) position in the horizontal plane. Single scatter interactions can be distinguished from multiple scatters, as the latter feature more than one S2 signal. 
Due to the different specific energy loss along the track, the S2/S1 ratio is different for ER, where the interaction is with atomic electrons (from $\gamma$, $\beta$ and neutrino backgrounds), and for NR, where the interaction is with the nucleus itself (from WIMPs, neutrons or neutrinos). 
WIMPs are expected to produce a single interaction in the active region, uniformly distributed in the volume, and of the NR type. 
We thus select NRs using the S2/S1 ratio to discriminate the WIMP signal from the ER background and requiring a single scatter interaction occurring in the central part of the active volume, since the background from the detector components is larger in the outer layers of the TPC. 

\subsection{Detector model}
The MC simulation of the XENON1T experiment is developed using the GEANT4 toolkit \cite{G4}. 
The model features all components of significant mass or impact for the photon collection efficiency and was built according to the CAD construction drawings. 
A rendered model of the XENON1T detector, as it is implemented in the MC simulation, is shown in figure \ref{fi:detector}.

The total amount of about $3.5$~t of LXe is contained in a vacuum-insulated double-wall cryostat made of $5$~mm thick low radioactivity stainless steel (SS). The dimensions of the inner cryostat are chosen to host the XENON1T TPC, while the outer vessel is sufficiently large to host also the future upgraded version of the experiment, XENONnT. Both vessels are composed of a central cylindrical part, closed by a dome on each side. The top domes are attached to the cylindrical part through two flanges, each $45$~mm thick. The cryostats are connected to the XENON1T cryogenic system via a long double-wall vacuum-insulated pipe, starting from a central port in the top domes.

The target inside the TPC consists of about $2$~t of LXe, constrained laterally by an approximately cylindrical structure of $24$ interlocking polytetrafluoroethylene (PTFE) panels with a radius of $\sim 480$~mm. The target volume is viewed by two arrays of PMTs. One is directly immersed in LXe at the bottom of the TPC, and consists of $121$ PMTs in a compact hexagonal structure, to maximize the light collection efficiency. The second is placed in GXe above the target volume, and is made of $127$ PMTs arranged in concentric rings to improve the radial position reconstruction. 
While the bottom array fits into the TPC radius, the radius of the top array is $\sim 40$~mm larger, to achieve a good resolution in the position reconstruction also at the edge of the TPC.
The space between the PMTs is covered with PTFE to reflect the vacuum ultraviolet (VUV) light \cite{PTFE-VUV} and ensure a high light collection efficiency. 
The structure of the TPC is sustained on the outside region through PTFE pillars. 
Additional PTFE and copper disks support the two PMT arrays.

The PMTs are $3"$ Hamamatsu R11410-21, specifically developed for XENON1T, and chosen for their high quantum efficiency ($35\%$ on average) and low radioactivity, which was evaluated through a screening campaign performed by the XENON1T collaboration \cite{PMTscreening}. Their characteristics and performance are described in more detail in \cite{PMTperformances1, PMTperformances2}. The main components of the PMT, reproduced in the GEANT4 model, are a cobalt-free body made of Kovar, a quartz window and a ceramic stem, with some smaller parts made of SS. We did not include in the model some internal components as the dynodes and the getter, nor the tiny ones (as the pins).
The voltage divider circuit is mounted on a base made of {\it Cirlex}, modeled as a simple thin disk.

The electric fields in the TPC are generated through electrodes made of SS meshes or wires stretched onto SS rings. There are two electrodes on the bottom of the TPC:  the cathode and a ground mesh to screen the bottom PMT array from the high electric field. At the liquid/gas interface there is a stack of two electrodes: the grounded gate electrode slightly below the surface and the anode right above it, separated by $5$~mm, creating the proportional amplification region. Another ground mesh above the anode is used to protect the top PMTs. The distance between the cathode and the gate meshes, which defines the active region 
where all generated light and charge signals can be detected, is $967$~mm, taking into account the $1.5\%$ shrinkage of PTFE components in the cooling from room to LXe temperature. A stack of $74$ field shaping rings, made of OFHC copper and placed just outside the PTFE lateral panels, ensures the uniformity of the electric field within the TPC. 

The liquid level in the proportional amplification region is adjusted between the gate and the anode electrode, and kept constant by using the concept of a diving bell with an adjustable overflow tube coupled to a linear motion feedthrough \cite{xe100-instrument}. The bell closing the GXe region is made of SS,  $5$~mm thick on the top and $3$~mm in the lateral part. This solution has the additional advantage that the LXe outside the bell can rise above the top PMT array, such that there is a $5$~cm layer of LXe above the bell and all around the TPC (outside of the field cage, between the TPC and cryostat walls) and $3$~cm of LXe below the basis of the bottom PMT array. 
This LXe layer acts as a passive shield, reducing the external backgrounds.

The region inside the inner cryostat, below the lower PMT array, was supposed to host an empty container made of $5$~mm thick SS along the cryostat wall and a $10$~mm flat cap on top. The aim was to reduce the total amount of xenon and to be used as a reservoir. During the detector installation this reservoir was not used, because of the availability of xenon. However it is still present in the background study presented in this work, leading to a conservative estimation.

A summary of the main materials used in the MC model is presented in Table \ref{ta:screening}, together with their measured radioactive contaminations (which will be described in section \ref{se:screening}).

We also modeled the other large components shown in figure \ref{fi:detector} (left), namely the pipes, the structure which sustains the cryostat, the water and the SS tank. However, given the larger distance from the TPC, their contribution to the background has been found to be negligible, therefore they are ignored in the following study. The gamma and neutron backgrounds from environmental radioactivity are also reduced to a negligible level due to the large water shield surrounding the detector \cite{MS-IDM2010}.


\landscape
\begin{table}[h]
\begin{center}\resizebox*{1.5\textwidth}{!}{ 
\begin{tabular}{|ccccccccccccc|}
\hline
\rule[-4mm]{0mm}{1cm}
\bf \Large ID & \bf \Large Component & \bf \Large Material &\bf \Large Quantity & \bf \Large Unit   & \multicolumn{8}{c|}{\Large \textbf{Contamination} $[\textrm{mBq/unit}]$} \\
\hline
\rule[-4mm]{0mm}{1cm}
\bf    & \bf    & \bf    & \bf   &\bf    &\bf \Large ${}^{238}\textrm{U}$ &\bf \Large ${}^{235}\textrm{U}$ &\bf \Large ${}^{226}\textrm{Ra}$ &\bf \Large ${}^{232}\textrm{Th}$ &\bf \Large ${}^{228}\textrm{Th}$ &\bf \Large ${}^{60}\textrm{Co}$ &\bf \Large ${}^{40}\textrm{K}$ &\bf \Large ${}^{137}\textrm{Cs}$ \\ 
\hline
\rule[-4mm]{0mm}{1cm}
1 & \bf  Cryostat Shells  & SS & 870 & kg    & $2.4 \pm 0.7$ & $(1.1 \pm 0.3) \cdot 10^{-1}$  & $< 6.4 \cdot  10^{-1}$ & $(2.1 \pm 0.6) \cdot 10^{-1}$ & $< 3.6  \cdot  10^{-1}$ & $9.7 \pm 0.8$ & $< 2.7$ & $< 6.4  \cdot  10^{-1}$ \\
\rule[-4mm]{0mm}{1cm}
2 & \bf  Cryostat Flanges & SS & 560 & kg           & $1.4 \pm 0.4$ &  $(6 \pm 2)\cdot 10^{-2}$ & $< 4.0$ & $(2.1 \pm 0.6) \cdot 10^{-1}$ & $4.5 \pm 0.6$ & $37.3 \pm 0.9$ & $< 5.6$ & $< 1.5$ \\
\rule[-4mm]{0mm}{1cm}
3 & \bf  Reservoir   & SS & 90  & kg        & $11 \pm 3$ & $(5 \pm 2) \cdot 10^{-1}$ & $1.2 \pm 0.3$ & $1.2 \pm 0.4$ & $2.0 \pm 0.4$ & $5.5 \pm 0.5$ & $< 1.3$ & $< 5.8 \cdot 10^{-1}$ \\
\rule[-4mm]{0mm}{1cm}
4 & \bf  TPC Panels $^{(1)}$    & PTFE & 92 & kg   & $< 2.5 \cdot 10^{-1}$ & $< 1.1 \cdot 10^{-2}$ & $< 1.2 \cdot 10^{-1}$ & $< 4.1 \cdot 10^{-2}$ & $< 6.5 \cdot 10^{-2}$ & $< 2.7 \cdot 10^{-2}$ & $< 3.4 \cdot 10^{-1}$ & $(1.7 \pm 0.3) \cdot 10^{-1}$\\
\rule[-4mm]{0mm}{1cm}
5 & \bf  TPC Plates  $^{(2)}$   & Cu & 184 & kg   & $< 1.2$ & $< 5.5 \cdot 10^{-1}$  &  $< 3.3 \cdot 10^{-2}$ & $< 4.3 \cdot 10^{-2}$  & $< 3.4 \cdot 10^{-2}$ & $0.10 \pm 0.01$ & $< 2.8 \cdot 10^{-1}$ & $< 1.6 \cdot 10^{-2}$  \\
\rule[-4mm]{0mm}{1cm}
6 & \bf  Bell and Rings $^{(3)}$   & SS & 80 & kg  & $2.4 \pm 0.7$ & $(1.1 \pm 0.3) \cdot 10^{-1}$  & $< 6.4 \cdot 10^{-1}$ & $(2.1 \pm 0.6) \cdot 10^{-1}$ & $< 3.6 \cdot 10^{-1}$ & $9.7 \pm 0.8$ & $< 2.7$ & $< 6.4 \cdot 10^{-1}$ \\
\rule[-4mm]{0mm}{1cm}
7 & \bf  PMT Stem &  Al$_2$O$_3$ & 248 & PMT  & $2.4 \pm 0.4$ & $(1.1 \pm 0.2) \cdot 10^{-1}$ & $(2.6 \pm 0.2) \cdot 10^{-1}$ & $(2.3 \pm 0.3) \cdot 10^{-1}$ & $(1.1 \pm 0.2) \cdot 10^{-1}$ & $ < 1.8 \cdot 10^{-2}$ & $1.1 \pm 0.2$ & $< 2.2 \cdot 10^{-2}$ \\
\rule[-4mm]{0mm}{1cm}
8 & \bf  PMT Window & Quartz & 248  & PMT  & $ < 1.2$  & $ < 2.4 \cdot 10^{-2}$ & $(6.5 \pm 0.7) \cdot 10^{-2}$ & $ < 2.9 \cdot 10^{-2}$ & $< 2.5 \cdot 10^{-2}$ & $ < 6.7 \cdot 10^{-3}$ & $< 1.5 \cdot 10^{-2}$ & $< 6.8 \cdot 10^{-3}$\\
\rule[-4mm]{0mm}{1cm}
9 & \bf  PMT SS & SS & 248  & PMT  & $(2.6 \pm 0.8) \cdot 10^{-1}$ & $(1.1 \pm 0.4) \cdot 10^{-2}$  & $< 6.5 \cdot 10^{-2}$ & $< 3.9 \cdot 10^{-2}$ & $< 5.0 \cdot 10^{-2}$ & $(8.0 \pm 0.7) \cdot 10^{-2}$ & $< 1.6 \cdot 10^{-1}$ & $< 1.9 \cdot 10^{-2}$\\
\rule[-4mm]{0mm}{1cm}
10 & \bf  PMT Body &  Kovar & 248  & PMT  &  $< 1.4 \cdot 10^{-1}$ & $< 6.4 \cdot 10^{-3}$ & $< 3.1 \cdot 10^{-1}$ & $< 4.9 \cdot 10^{-2}$ & $< 3.7 \cdot 10^{-1}$ & $(3.2 \pm 0.3) \cdot 10^{-1}$ & $< 1.1$ & $< 1.2 \cdot 10^{-1}$\\
\rule[-4mm]{0mm}{1cm}
11 & \bf  PMT Bases & Cirlex & 248  & PMT  & $(8.2 \pm 0.3) \cdot 10^{-1}$ & $(7.1 \pm 1.6) \cdot 10^{-2}$ & $(3.2 \pm 0.2) \cdot 10^{-1}$ & $(2.0 \pm 0.3) \cdot 10^{-1}$ & $(1.53 \pm 0.13) \cdot 10^{-1}$ & $< 5.2 \cdot 10^{-3}$ & $(3.6 \pm 0.8) \cdot 10^{-1}$ & $< 9.8 \cdot 10^{-3}$\\
\rule[-4mm]{0mm}{1cm}
12 & \bf  Whole PMT              & - & 248     & PMT       & $8 \pm 2$ &  $(3.6 \pm 0.8) \cdot 10^{-1}$ & $(5 \pm 1) \cdot10^{-1}$ & $(5 \pm 1) \cdot 10^{-1}$ & $(5.0 \pm 0.6) \cdot 10^{-1}$ & $(7.1 \pm 0.3) \cdot 10^{-1}$ & $13 \pm 2$ & $< 1.8 \cdot 10^{-1}$ \\

\hline 
\end{tabular} }
\end{center}
\caption{Summary of the contaminations of the materials considered in the XENON1T Monte Carlo simulation.\newline
$^{(1)}$ includes all the PTFE components present in the TPC walls, rings and pillars, the plates of the two support structures of the PMTs, and a trapezoidal section ring on top of the TPC; \newline
$^{(2)}$ field shaping rings and support structure of the two PMT arrays; \newline 
$^{(3)}$ top and lateral part of the bell; support rings of the five electrodes.}
\label{ta:screening}
\end{table}
\endlandscape

\subsection{Radioactive contamination in detector materials}
\label{se:screening}
The selection of materials for the detector components is based on an extensive radioactivity screening campaign, using mainly two complementary techniques and dedicated measurements: germanium (Ge) detectors and mass spectrometry. For the gamma spectrometry with Ge, XENON has access to the most sensitive screening facilities: the Gator \cite{GATOR} and GeMPIs \cite{Heusser2006495} detectors, placed underground at LNGS, and GIOVE \cite{Heusser:2015ifa} in Heidelberg. Ge detectors are sensitive to most of the radiogenic nuclides relevant for the ER and NR background: $^{40}\rm{K}$, $^{60}\rm{Co}$, $^{137}\rm{Cs}$, those in the $^{232}\rm{Th}$ chain and in the second part of the $^{238}\rm{U}$ chain,  $^{226}\rm{Ra}$ and its daughters. The main exception is the first part of the $^{238}\rm{U}$ chain, where very few and low energy gammas are emitted. However, the estimation of the activity of this part of the chain is very important, since it is responsible for the production of neutrons from spontaneous fission, as discussed in section \ref{S:Nr}.  For this reason, we also used mass spectrometry techniques to directly count the amount of primordial nuclides ($^{238}\rm{U}$,  $^{232}\rm{Th}$). For some materials, we also inferred the $^{238}\rm{U}$ abundance, using the measurement of $^{235}\rm{U}$ with gamma spectroscopy, and scaling for their natural abundances.

In table \ref{ta:screening}, we report the contaminations for all the materials considered in the MC model and used in the prediction of the ER and NR background, 
presented in sections \ref{S:Em} and \ref{S:Nr}. 
As for XENON100 \cite{xe100-nrbkg}, to have a better characterization of the induced background, we considered in detail the disequilibrium in the $^{238}\rm{U}$ and  $^{232}\rm{Th}$ chains, splitting them in two parts: 
 \begin{itemize}
\item $^{238}\rm{U} \: \rightarrow \:   ^{230}\rm{Th}$  and $^{226}\rm{Ra} \:  \rightarrow \:  ^{206}\rm{Pb}$, for the   $^{238}\rm{U}$ chain,
\item  $^{232}\rm{Th} \:  \rightarrow \:   ^{228}\rm{Ac}$ and $^{228}\rm{Th} \:  \rightarrow \:   ^{208}\rm{Pb}$,  for the  $^{232}\rm{Th} $ chain.
 \end{itemize}
 Indeed, we observe disequilibrium for instance in the SS of the cryostat and in some of the PMT components.

The materials constituting the TPC come from various batches, and thus present different contaminations.
In row 4 to 6 of table \ref{ta:screening} we report only the contamination of the batch which contains the largest mass of PTFE, copper and SS, respectively.
For the estimation of the contamination of the PMTs, we followed two strategies: for the ER background we used the measurement in row $12$, obtained by measuring the whole PMT with the Ge detectors. For the estimation of the NR background, since the neutron yield depends also on the particular material, we considered the measurement of the raw PMT components (rows $7$--$10$). More details on the measurement of the contamination of the PMTs and their constituents can be found in \cite{PMTscreening}.
A whole description of the screening campaign, with additional results and details, will be presented in a dedicated article \cite{xenon1t-screening}.

When only upper limits were available, we treated them as detection values or set them to zero, to obtain the maximum and the minimum prediction of the background rate, respectively.

\subsection{Physics list in GEANT4}
For the MC simulation we used version 9.5-patch01 of the GEANT4 toolkit.
Radioactive decays of nuclei are simulated using the {\it G4RadioactiveDecay} process, featuring $\alpha$, $\beta^{+}$, $\beta^{-}$ decay and electron capture. Half lives, nuclear level structure, decay branching ratios, and energies of decay processes are data-driven, taken from the Evaluated Nuclear Structure Data Files (ENSDF) \cite{ENSDF}.
If the daughter of a nuclear decay is an isomer, prompt de-excitation is managed through the {\it G4PhotonEvaporation} process.
For gamma and electron interactions, we chose the {\it Livermore} physics list, particularly suited to describe the electromagnetic interactions of low energy particles. For low energy neutrons ($<$20~MeV) we use the {\it High Precision} physics list where the elastic, inelastic and capture processes are described in detail, 
using the neutron data files G4NDL 3.13 with thermal cross sections, which are
based on the ENDF/B-VI/B-VII databases \cite{ENDF}.

In GEANT4 the tracking of the various particles is divided into steps, whose length is automatically chosen according to the type and energy of the particle, and the characteristics of the medium.
For each step of all the particles inside the LXe target, we record the position, time, deposited energy, particle type and the process responsible for the energy loss.

\section{Electronic recoil background}
\label{S:Em}

In XENON1T, we distinguish ERs from the expected signal of NRs based on their different S2/S1 ratio. The typical rejection efficiency achieved in XENON100 is of the order of $99.5\%$ at $50\%$ signal acceptance~\cite{xe100-run10}. However, potential statistical leakage of ER events into the NR region can mimic a WIMP signal. Thus, we considered all the relevant sources of ER background: radioactive contamination of the detector materials, radioactive isotopes intrinsic to the LXe ($^{222}\rm{Rn}$ and its daughters, $^{85}\rm{Kr}$, and $^{136}\rm{Xe}$ double-beta decays) and solar neutrinos scattering off electrons.

We select the background events requiring a single scatter interaction in the TPC, occurring in the fiducial volume (FV, the most internal region of the LXe target), and in the low energy range.

To determine the single scatter, we adopted the following strategy: we convert the deposited energy into the number
of produced electrons, following the procedure detailed in section \ref{S:Conv}, and we apply a selection
similar to the one currently used in the XENON100 analysis \cite{xe100-analysis}. Namely, we require the second largest S2 to be smaller than 5 electrons.
The capability to distinguish two scatters is related also to the width of the S2 signal peaks in the time domain, and to the peak separation efficiency of the S2 peak finder algorithm. Based on the XENON100 performance~\cite{xe100-instrument}, a multiple scatter event is misidentified as a single scatter if the interactions occur within $3$~mm in the vertical direction. 
In this analysis, we conservatively did not use the (x, y) information to separate two scatters occurring within the same horizontal plane.

The reference $1$~t FV is defined as a super-ellipsoid of third degree, centered in the middle of the active region, with radius and semi-height equal to $40$~cm, such that a minimum distance of $4$~cm to the borders of the TPC is guaranteed. However, we also considered other FVs in our analysis modifying both radius and semi-height by the same factor.

With the typical velocities from the standard galactic halo model \cite{MCSmith}, WIMPs are expected to produce NRs in xenon with kinetic energies mostly below $50$~keV, with a strong dependence on the WIMP mass.
Considering the different light response of LXe to ER interactions (a detailed discussion will follow in section \ref{S:Conv}), 
this NR energy corresponds to an ER energy of about $12$~keV. To avoid fake events generated by accidental coincidence of PMT dark counts, we use a lower energy threshold of $1$~keV.

\subsection{Radioactivity from detector components}
For each of the components listed in table \ref{ta:screening}, we generated the decays of all the isotopes, confining their origin uniformly inside their volume.  
The background events are due to $\gamma$-rays that reach the internal volume of the active region, producing a low energy Compton scatter and exiting the detector without other interactions. 
We generated about $10^9$ decays for each detector component $i$ and radioactive isotope $j$. 
By using the selection criteria introduced above, we obtain the number of surviving events, $N_{i,j}$. 
The differential background rate, $R_{i,j}$, is calculated as:
\begin{equation}
 R_{i,j} = N_{i,j} / (T_{i,j} \cdot M_\mathrm{FV} \cdot \Delta E),
 \end{equation}
where $M_\mathrm{FV}$ is the fiducial mass considered, $\Delta E$ is the energy range, and $T_{i,j}$ is the effective livetime given by:
\begin{equation}
T_{i,j} = N^{G}_{i,j} / ( M_{i} \cdot A_{i,j} ), 
\end{equation}
where $N^{G}_{i,j}$ is the number of generated events in each MC simulation,
$M_{i}$ and $A_{i,j}$  are the masses and the specific activities reported in table \ref{ta:screening}. 

\begin{figure}[t!]
\centering\includegraphics[width=0.9\linewidth, keepaspectratio]{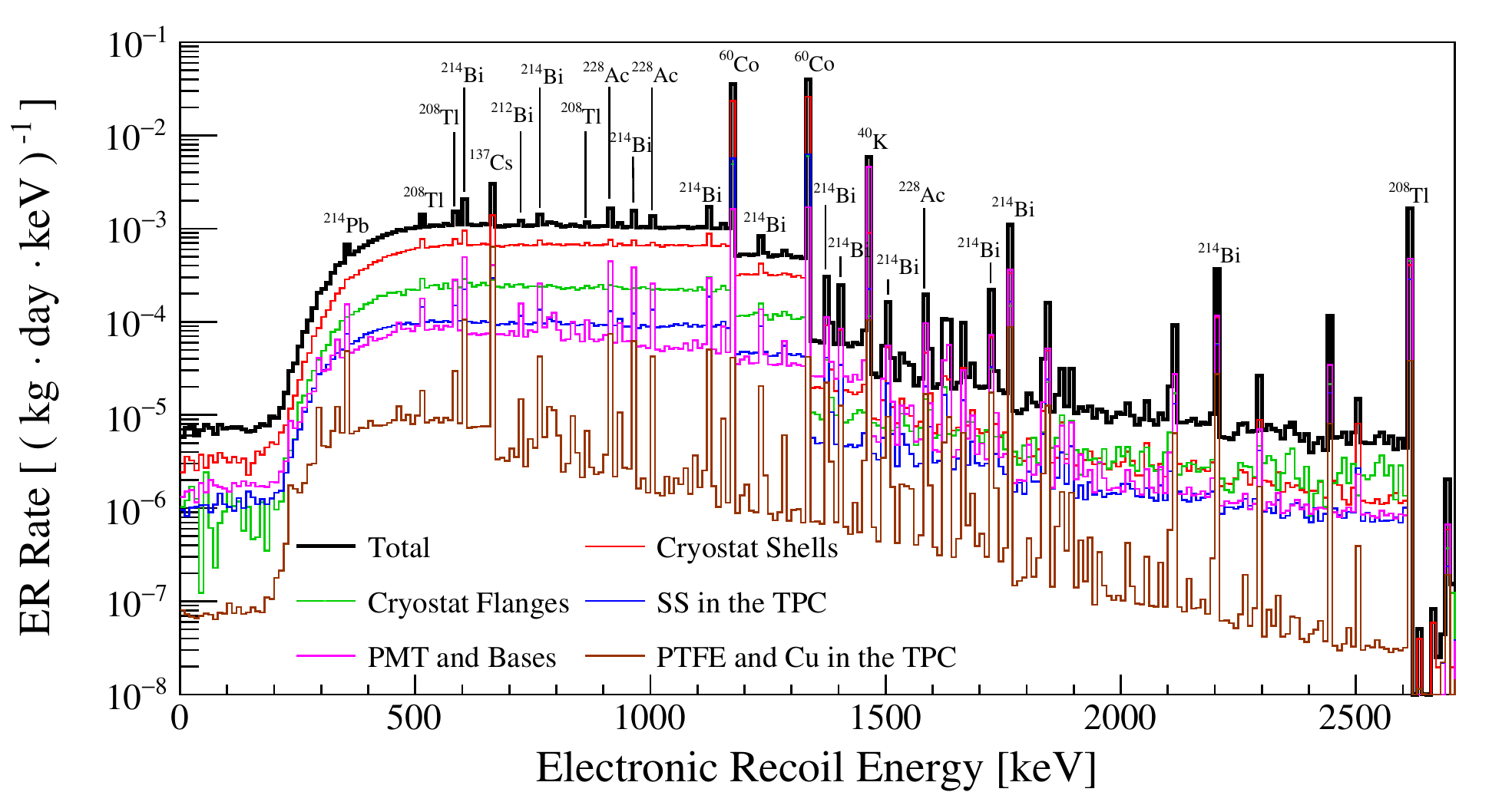} 
\caption{Energy spectrum in $1$~t FV of the total ER background from the detector materials (black), and the separate contributions from the various components (colors).}
\label{fi:er-spectrum}
\end{figure}

The recoil energy spectrum \footnote{The smearing due to the energy resolution of the detector is not applied to the results shown throughout section \ref{S:Em} and \ref{S:Nr}} of background events in the $1$~t FV is shown in figure \ref{fi:er-spectrum}. 
At low energies, below $200$~keV, the spectrum is generated by Compton scatter processes and it is almost flat, while at higher energies the various photo-absorption peaks are visible.
The total background rate from materials in the ($1$, $12$)~keV energy range is $(7.3 \pm 0.7) \cdot 10^{-6}$~\dru, corresponding to $(30 \pm 3)$~y$^{-1}$ in 1~t FV. The uncertainty includes those from the measurements of the material contaminations, the statistical uncertainty in $N_{i,j}$ and a $10\%$ systematic uncertainty accounting for 
the potential differences in modeling the actual geometry of the detector in the MC (e.g. the use of the reservoir, the level of LXe above the diving bell).
The highest contribution, $61\%$ of the total ER background from materials, comes from the SS of the cryostat (shells and flanges), mainly from its $^{60}\rm{Co}$ contamination. The PMTs with their bases contribute $23\%$, while $15\%$ is due to the other SS components inside the TPC (reservoir, diving bell, electrode rings). The ER background contribution from the PTFE and copper parts is about $1\%$.


\begin{figure}[t]
\centering

\mbox{
\begin{minipage}[t]{0.49\textwidth}
\centering

\includegraphics[width=1.\linewidth]{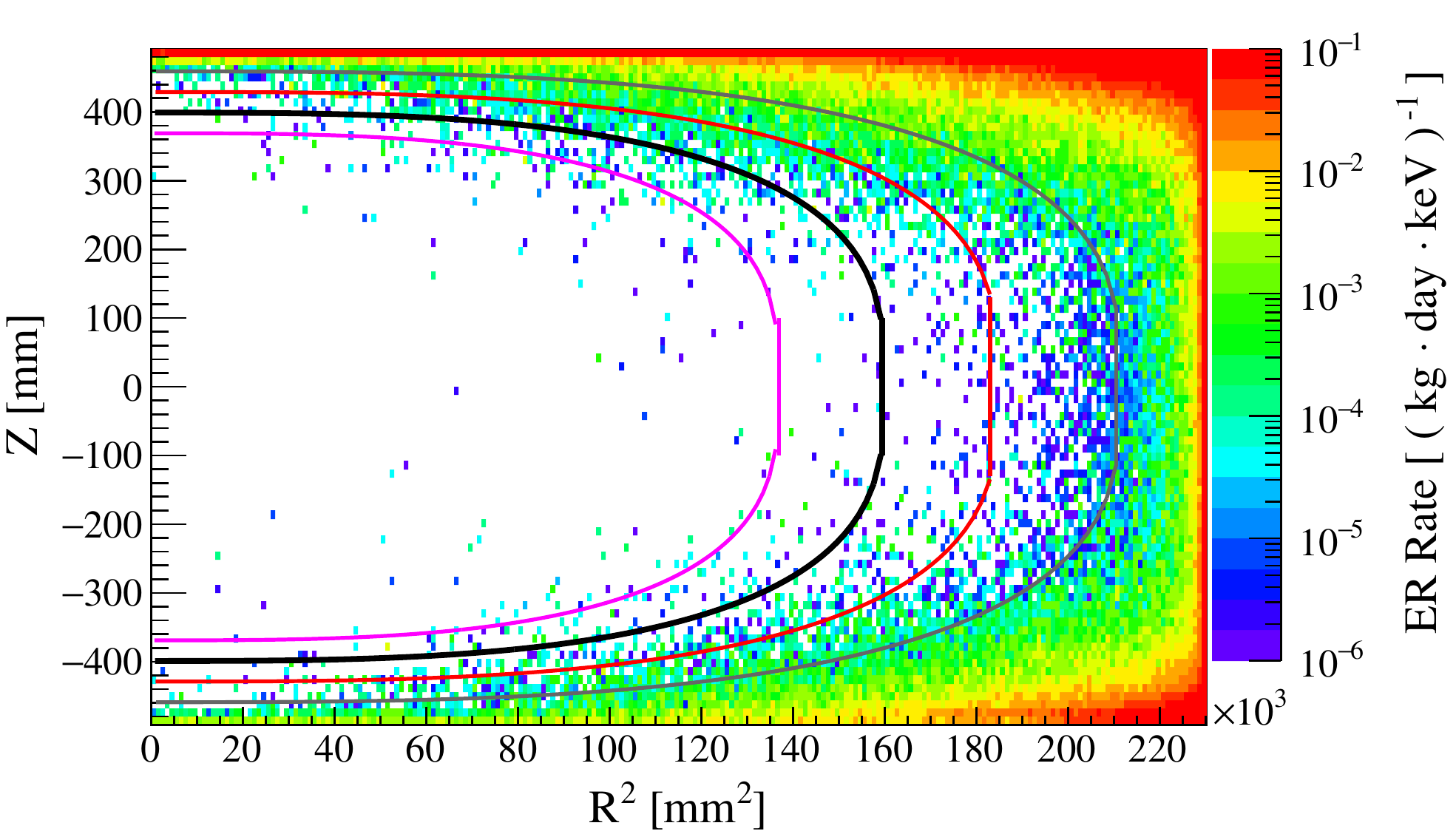}
\caption{Spatial distribution of the ER background events from the detector materials inside the active LXe volume, in the ($1$, $12$)~keV energy range. The thick black line indicates the reference $1$~t super-ellipsoid fiducial volume. With the purple, red and brown lines, we indicate the FVs corresponding to $800$~kg, $1250$~kg and $1530$~kg, respectively. The white regions present a background rate smaller than $1 \cdot 10^{-6}$~\dru.}
\label{fi:er-zr}

\end{minipage}
\hspace{0.02\textwidth}
\begin{minipage}[t]{0.49\textwidth}
\centering

\includegraphics[width=.95\linewidth]{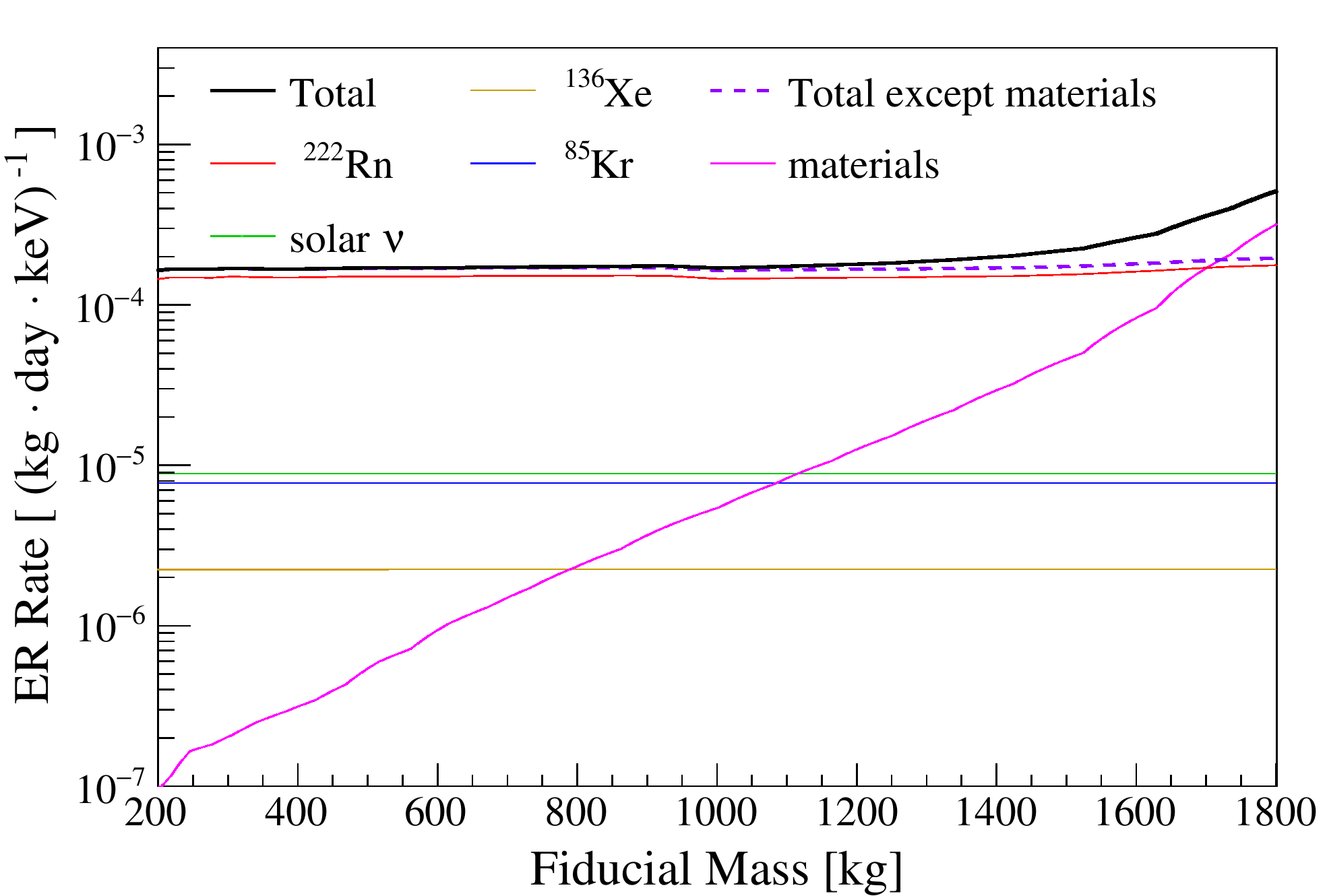}
\caption{The total ER background rate as a function of the fiducial mass (black line), together with the separate contributions from the detector components (purple), $10$~$\mu$Bq/kg of $^{222}\rm{Rn}$ (red), $0.2$~ppt of $^\mathrm{nat}\rm{Kr}$ (blue), solar neutrinos (green) and  $^{136}\rm{Xe}$ double-beta decay (brown). With the dashed violet line we show the sum of the background sources uniformly distributed inside the LXe volume. The rate is averaged over the energy range ($1$, $12$)~keV.}
\label{fi:er-bkgvsmass}

\end{minipage}
}

\end{figure}

A lower bound in the background level can be evaluated by setting to zero all the contaminations that are reported as upper limits in table \ref{ta:screening}. In this case the total background rate is $(6.7 \pm 0.7) \cdot 10^{-6}$~\dru, corresponding to $(27 \pm 3)$~y$^{-1}$  in $1$~t FV, about $10\%$ smaller than the previous estimate.

The spatial distribution of the background events inside the whole active volume, in the energy range ($1$, $12$)~keV, is shown in figure \ref{fi:er-zr} together with some fiducial volumes (corresponding to $800$, $1000$, $1250$ and $1530$ kg).
The background rate as a function of the fiducial mass is shown in figure \ref{fi:er-bkgvsmass}.

\subsection{$\bf ^{222}\rm{Rn}$}
In XENON1T, the main intrinsic source of background in LXe comes from the decays of $^{222}\rm{Rn}$ daughters. Being part of the $^{238}\rm{U}$ decay chain, $^{222}\rm{Rn}$ can emanate from the components of the detector and the gas system, or diffuse through the vacuum seals. Due to its relatively large half-life ($3.8$~days), it can homogeneously distribute inside the LXe volume. Considering $^{222}\rm{Rn}$ daughters, down to the long-lived $^{210}\rm{Pb}$, the most dangerous contribution comes from the $\beta$ decay of $^{214}\rm{Pb}$ to the ground state of $^{214}\rm{Bi}$, with an end-point energy of $1019$~keV, where no other radiation is emitted. According to GEANT4, version 10.0, the branching ratio for this channel is $10.9\%$ \footnote{Note that up to version 9.6 the branching ratio coded in GEANT4 was significantly smaller: $6.3\%$.}. However, especially if the decay occurs close to the borders of the active region, decays to other energy levels are also potentially dangerous since the accompanying $\gamma$ can exit the detector undetected. This is responsible for the slightly higher background rate from $^{222}\rm{Rn}$ seen at larger fiducial masses in figure \ref{fi:er-bkgvsmass}. Given the increased target mass in XENON1T, this effect is less relevant than what was observed in XENON100 \cite{xe100-erbkg}.
The only other $\beta$ emitter in the chain ($^{214}\rm{Bi}$), also a potential source of background, can be easily removed by looking at the time correlation with the $\alpha$ decay of its daughter, $^{214}\rm{Po}$, which occurs with a half-life of $164$~$\mu$s. Therefore, the contribution of $\beta$ decays of $^{214}\rm{Bi}$ is not considered in the background estimation.

To reduce the $^{222}\rm{Rn}$ concentration, XENON1T is built with materials selected for their low radon emanation. 
In addition, R\&D activities on cryogenic distillation are on-going to implement an online removal system, following the first assessment of the actual background in XENON1T.
In XENON100 we observed \cite{MarcWeberThesis} that after the $\alpha$-decay, the recoiling ions ($^{218}\rm{Po}$ and $^{214}\rm{Pb}$) are drifted towards the cathode. This results in a lower concentration of the $\beta$ emitters inside the fiducial volume, thus reducing the background with respect to the assumption of a uniform distribution in the LXe volume. However, 
given that the final reduction in rate due to this effect can be quantified
only after the first run of the experiment, in this study we conservatively neglect it.
From the measurements of the radon emanation of the materials in close contact with LXe (TPC and pipes), we estimate 
 a $^{222}\rm{Rn}$ contamination of $10$~$\mathrm{\mu}$Bq/kg, a reduction of a factor $\sim 5$ with respect to the current value achieved in XENON100. The predicted ER background in the ($1$, $12$)~keV energy region, in $1$~t FV, is $1.54 \cdot 10^{-4}$~\dru.
A $10\%$ systematic uncertainty is assumed for the background rate from $^{222}\rm{Rn}$, mainly due to the uncertainties in the branching ratio of the $^{214}\rm{Pb}$ decay to the ground state \cite{nucleide}.

The other radioactive noble gas, $^{220}\rm{Rn}$, has a lower probability to be diffused into the active LXe volume, due to its short half-life (56 s). Thus, assuming a concentration $< 0.1$~$\mathrm{\mu}$Bq/kg, the background induced by its daughters is considered negligible.

\subsection{$\bf ^{85}\rm{Kr}$}
\begin{figure}[t!]
\centering\includegraphics[width=0.49\linewidth]{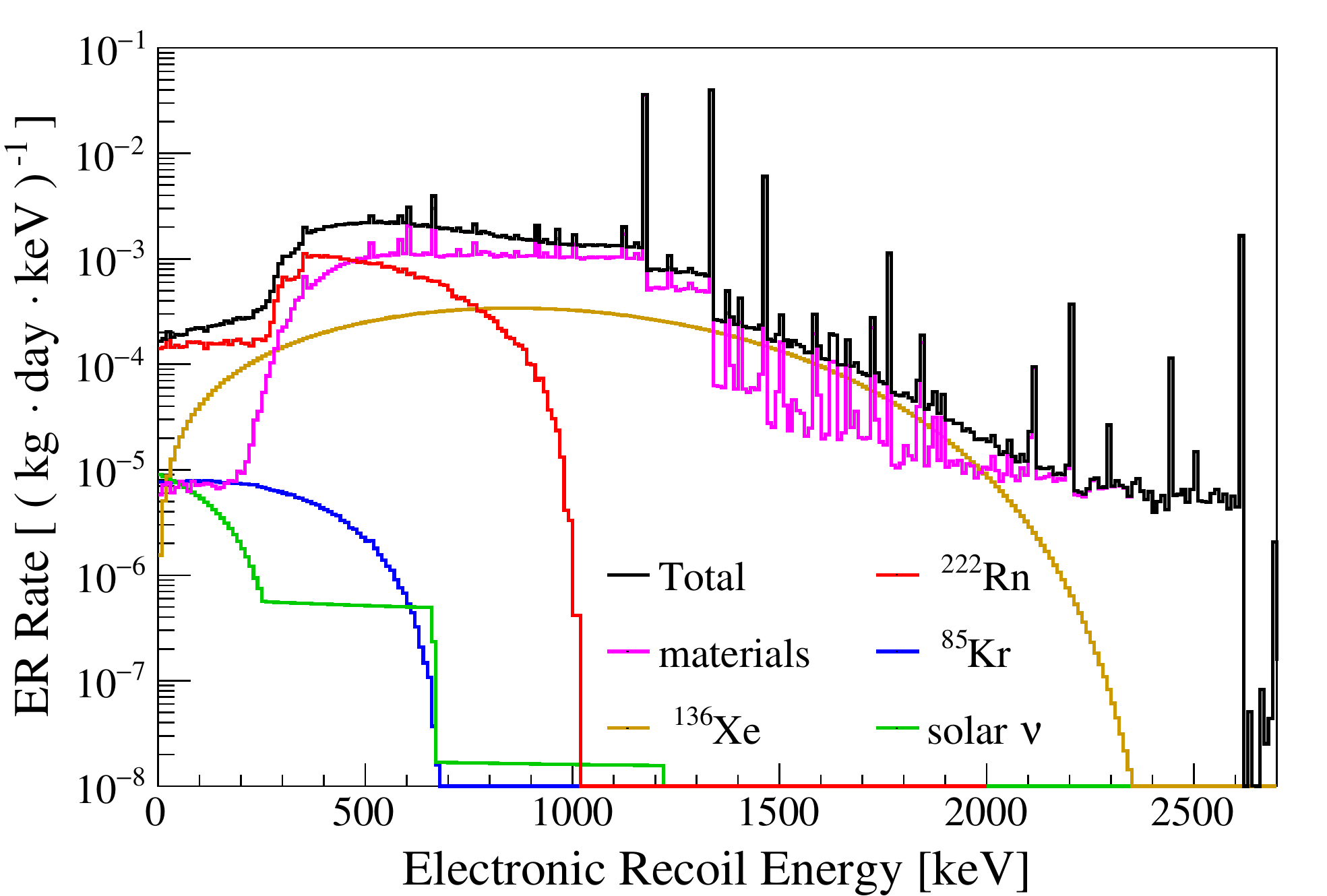}
\centering\includegraphics[width=0.49\linewidth]{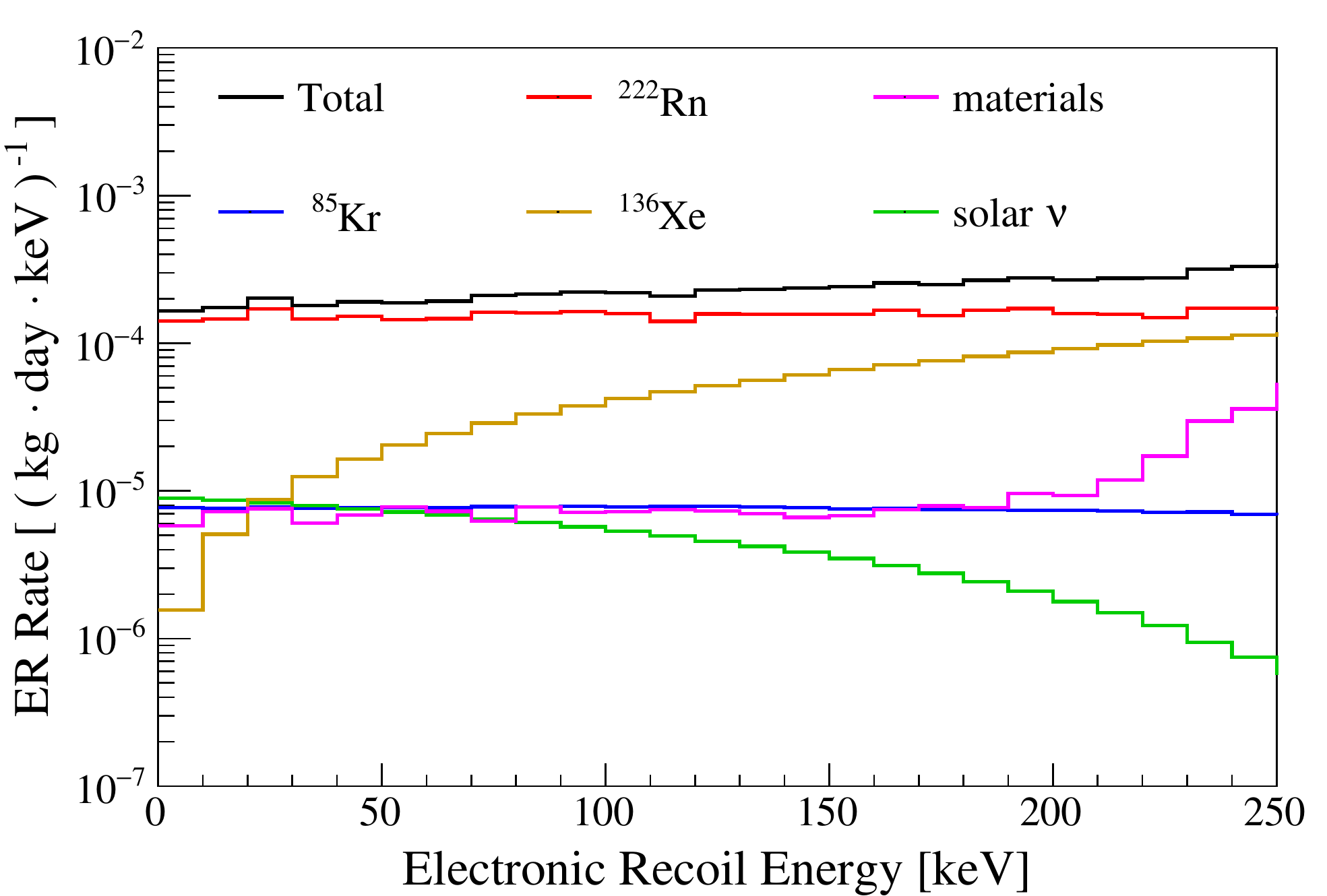}
\caption{Energy spectrum of the total ER background rate in the $1$~t fiducial volume (black), and the separate contributions from detector components (purple), $10$~$\mu$Bq/kg of $^{222}\rm{Rn}$ (red), $0.2$~ppt of $^\mathrm{nat}\rm{Kr}$ (blue), solar neutrinos (green) and  $^{136}\rm{Xe}$ double-beta decay (brown). The right plot shows the zoom at low energies.}
\label{fi:er-spectra}
\end{figure}

Xenon is extracted from the atmosphere with a typical $^\mathrm{nat}$Kr/Xe concentration at the ppm level. 
Natural krypton contains traces of the radioactive isotope $^{85}\rm{Kr}$, mainly produced by nuclear fission and released by nuclear fuel reprocessing plants and by nuclear weapon tests.
Its relative isotopic abundance in Europe has been determined by low level counting to be $2 \cdot 10^{-11}$ \cite{Sebastian}. $^{85}\rm{Kr}$ is a $\beta$ emitter with a half-life of $10.76$~y and an end-point energy of $687$~keV. The low energy tail of its $\beta$ spectrum, shown in figure \ref{fi:er-spectra} (blue line), extends into the region of interest for the WIMP search.
Due to the uniform distribution of Kr inside the LXe volume, it is not possible to reduce this source of background by fiducialization.

During the 225 live-days science run of XENON100, a  $^\mathrm{nat}$Kr/Xe concentration of $(19 \pm 4)$~ppt [mol/mol] has been achieved by processing the gas through cryogenic distillation \cite{xe100-run10}. Considering some samples drawn from the output of the XENON100 cryogenic distillation column, the krypton concentration was measured to be  $(0.95 \pm 0.16)$~ppt, the purest xenon target ever employed in a LXe particle detector \cite{Lindemann:2013kna}.
For XENON1T the goal is to further reduce the $^\mathrm{nat}$Kr/Xe concentration to $0.2$~ppt, using a new high through-put and high separation cryogenic distillation column \cite{Kr-column}. Sub-ppt concentrations have also already been achieved in tests of the cryogenic distillation column for XENON1T \cite{DominickThesis, StephanRosendahlPhD}.
 
The corresponding ER background in the low energy region  ($1$, $12$)~keV is $7.7 \cdot 10^{-6}$~\dru, very similar to the ER background from materials in the $1$~t FV. We adopt a $20\%$ uncertainty on the background rate from $^{85}\rm{Kr}$, mainly due to the uncertainty in the shape of the $\beta$ spectrum at low energies \cite{LRT2013}.

\subsection{$\bf ^{136}\rm{Xe}$ double-beta decay}
Natural xenon contains $8.9\%$ of $^{136}\rm{Xe}$, a two-neutrino double-beta emitter with a $Q$-value of $2458$~keV. The most recent measurement of its half-life 
is $2.17 \cdot 10^{21}$~y \cite{Exo}. The double-beta decay energy spectrum is obtained from the DECAY0 \cite{decay0}  code and is shown in figure \ref{fi:er-spectra} (brown line). At low energies, its parameterization is given by $(3.5 \cdot E / \mathrm{keV}) \cdot 10^{-7}$~\dru. The average background rate in the energy region ($1$, $12$)~keV is $2.3 \cdot 10^{-6}$~\dru. 
Following the discussion in \cite{Kotila:2012zza, IachelloWebSite}, we assume a  $15\%$ uncertainty for the low energy part of the $\bf ^{136}\rm{Xe}$ double-beta decay spectrum.

\subsection{Solar neutrinos}
Solar neutrinos scatter elastically off the electrons of the medium, producing ER in the low energy region. We considered neutrinos from all nuclear reactions in the Sun \cite{BP05, Serenelli:2011py}, taking into account neutrino oscillation $\nu_e \rightarrow \nu_{\mu, \tau}$ with an electron neutrino survival probability $P_{ee}=0.55$ \cite{PDG2014} and the reduced cross section for $\nu_{\mu, \tau}$. The resulting recoil energy spectrum is shown in figure \ref{fi:er-spectra} (green line), together with the other ER backgrounds. Most of the interactions ($92\%$) come from $pp$ neutrinos,  $^{7}\rm{Be}$ contributes with $7\%$, while {\it pep} and all the others sources contribute less than $1\%$. Below $100$~keV the differential rate is slightly decreasing, which we parameterize as $(9.155 - 0.036 \cdot E / \mathrm{keV}) \cdot 10^{-6}$~\dru. The average background rate in the energy region ($1$, $12$)~keV is $8.9 \cdot 10^{-6}$~\dru. 
This background source cannot be reduced and the only way to mitigate its impact is to improve the ER rejection \cite{Schumann:2015cpa}. 
We estimate a total $2\%$ uncertainty on the ER background events from solar neutrinos, obtained by combining the $\sim 1\%$ error in the flux from the $pp$ chain and $\sim 10\%$ in the $^{7}\rm{Be}$ \cite{Serenelli:2011py}, and adding the $\sim 2\%$ uncertainty in the oscillation parameter $\sin^2(2\theta_{12})$ \cite{PDG2014}. 

\subsection{Summary of ER backgrounds}
The ER background spectrum in $1$~t FV is summarized in figure \ref{fi:er-spectra}. The average rate of the total background in the ($1$, $12$)~keV range is $(1.80 \pm 0.15) \cdot 10^{-4}$~\dru. Most of the background ($\sim 85\%$) comes from $^{222}\rm{Rn}$, where we assumed a $10$~$\mathrm{\mu}$Bq/kg contamination. Solar neutrinos, $^{85}\rm{Kr}$ (assuming a $^\mathrm{nat}$Kr/Xe concentration of $0.2$~ppt), and ER from the materials contribute each with $(4-5) \%$. Even with the $8.9\%$ natural abundance of $^{136}\rm{Xe}$, its contribution to the total ER background is subdominant, less than $2\%$.  
The results are summarized in table \ref{ta:er-summary}, together with their uncertainties.
For energies larger than $500$~keV the radiation from materials becomes dominant. 

\begin{table}[t!]
\resizebox*{1.\textwidth}{!}
{
\centering
\begin{tabular}{|cccc|}
\hline
\textbf{Source} & \textbf{Background} $[\textrm{\dru}]$ & \textbf{Background} $[\textrm{y}^{-1}]$ & \textbf{Fraction} $[\%]$ \Tstrut\Bstrut \\ 
\hline
\bf Materials             & $(7.3 \pm 0.7)\cdot 10^{-6}$ & $30 \pm 3$ & ~4.1 \Tstrut \\
\bf ${}^{222}\textrm{Rn}$ & $(1.54 \pm 0.15)\cdot 10^{-4}$ & $620 \pm 60$ & 85.4 \\
\bf ${}^{85}\textrm{Kr}$  & $(7.7 \pm 1.5)\cdot 10^{-6}$ & $31 \pm 6$ & 4.3 \\
\bf ${}^{136}\textrm{Xe}$ & $(2.3 \pm 0.3)\cdot 10^{-6}$  & $~9 \pm 1$ & 1.4  \\
\bf Solar neutrinos       & $(8.9 \pm 0.2)\cdot 10^{-6}$ & $36 \pm 1$ & 4.9  \\
\hline
\textbf{Total}                 & $\mathbf{(1.80 \pm 0.15)\cdot 10^{-4}}$ & $\mathbf{720 \pm 60}$ & \textbf{100} \Tstrut\Bstrut \\
\hline
\end{tabular}
}
\caption{Summary of the predicted ER backgrounds in XENON1T, evaluated in $1$~t fiducial volume and in ($1$, $12$)~keV energy range. We assume $10$~$\mu$Bq/kg of $^{222}\rm{Rn}$, $0.2$~ppt of $^\mathrm{nat}\rm{Kr}$, and natural abundance of  $^{136}\rm{Xe}$.} 
\label{ta:er-summary}
\end{table}

The dependence of the background rates on the fiducial mass is shown in figure \ref{fi:er-bkgvsmass}: due to the large contribution of background from $^{222}\rm{Rn}$ which is almost uniformly distributed in the target volume, the one from the detector components becomes larger than the sum of the other backgrounds for fiducial masses larger than $1600$~kg.

\section{Nuclear recoil background}
\label{S:Nr}

Neutrons can produce NRs via elastic scattering off xenon nuclei. 
In addition, fast neutrons are more penetrating than 
$\gamma$-rays, their mean free path in LXe being on the order of tens of cm, thus 
they are more difficult to shield and their probability to have a single scatter in the LXe active volume is higher than for $\gamma$-rays.
They can generate a signal which is, on an event-by-event basis, indistinguishable from that of WIMPs. 
It is, therefore, crucial to minimize and accurately characterize this potentially dangerous background. 
The presence of isotopes of the primordial decay chains $^{238}\rm{U}$, $^{235}\rm{U}$ and $^{232}\rm{Th}$ 
 in the materials of the detector generates radiogenic neutrons in the MeV range through spontaneous fission (SF), mainly from $^{238}\rm{U}$,
and  ($\alpha$, n) reactions induced by the various $\alpha$ particles emitted along the decay chains. Additionally, cosmogenic neutrons with energies
extending to tens of GeV are produced by muons along their path through the rock into the underground laboratory and through the materials that surround the detector.

Astrophysical neutrinos, in particular those produced in the $^{8}\rm{B}$ decay in the Sun,
also contribute to the NR background as they can scatter coherently off the xenon nuclei, through coherent neutrino-nucleus scattering (CNNS).
In a detector where it is not possible to measure the direction of the recoil track, this is an irreducible background because it is a single scatter NR, uniformly distributed in the active volume of the TPC, defining the ultimate limitation to WIMP direct detection experiments \cite{Billard}. See \cite{Grothaus:2014hja} for a  review on the potential of future directional detectors. 


\subsection{Radiogenic neutrons from detector components}
\begin{figure}[t!]
\includegraphics[width=0.5\linewidth]{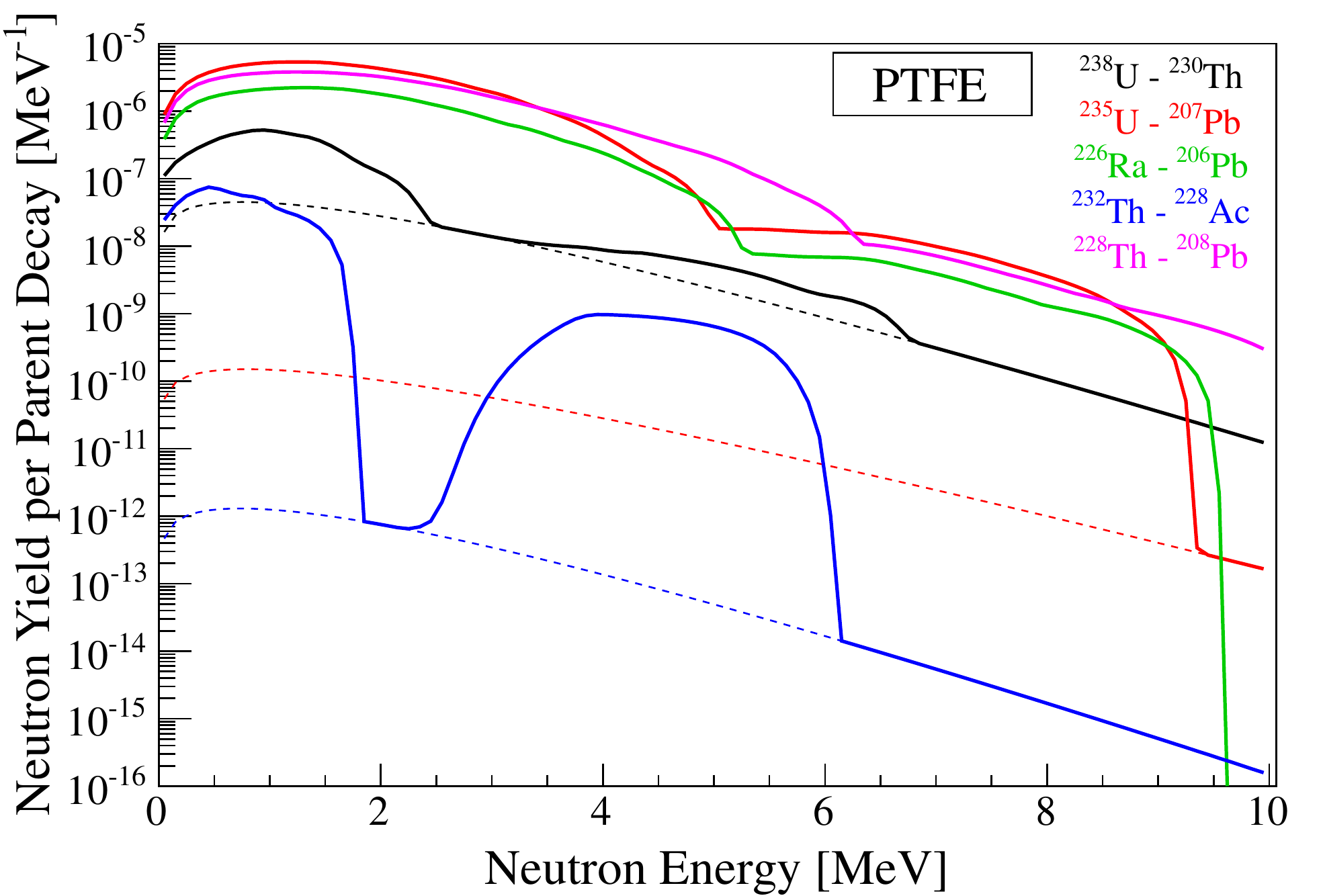}
\includegraphics[width=0.5\linewidth]{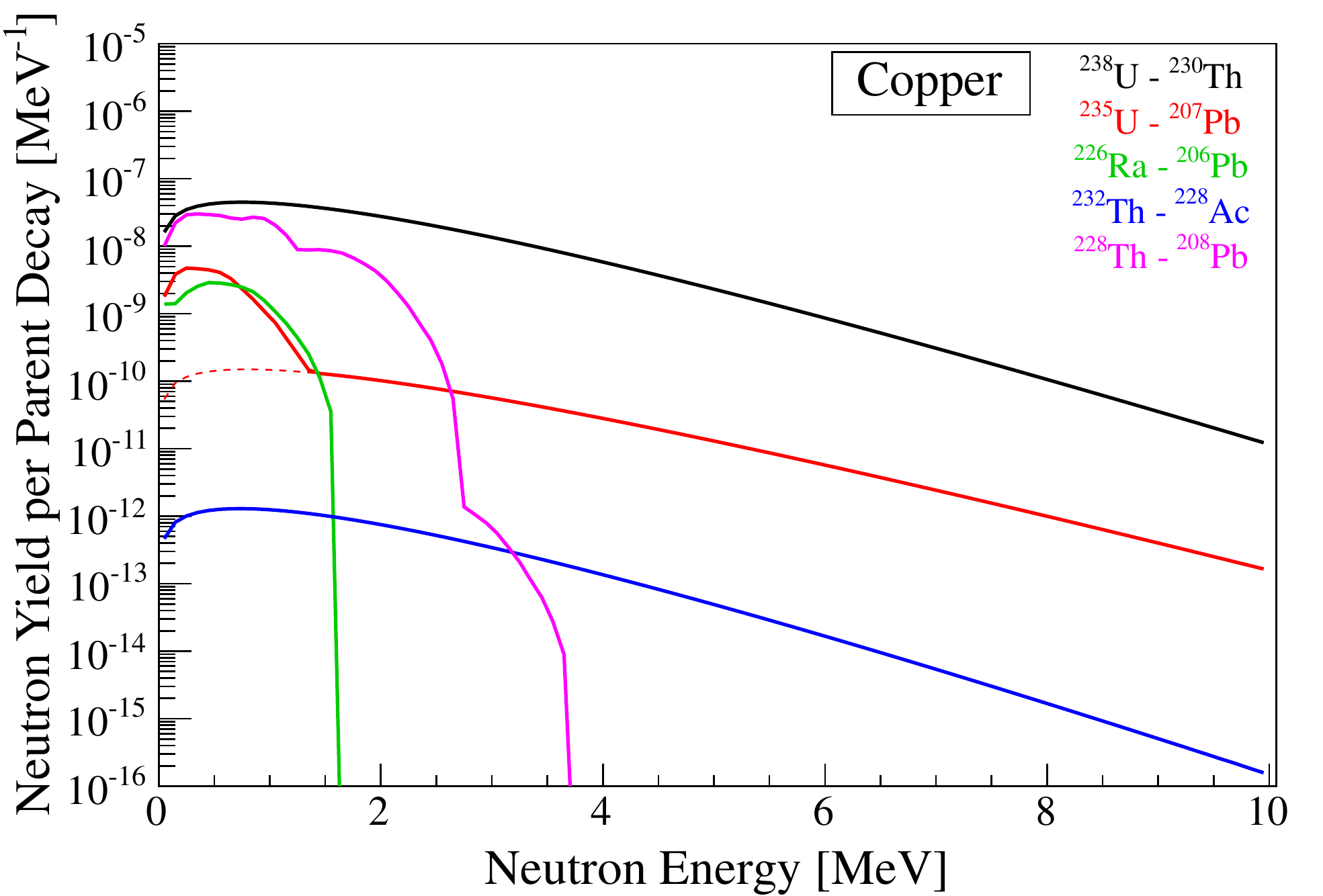}
\caption{Differential yield of radiogenic neutrons in PTFE (left) and copper (right), normalized per each decay of the parent nucleus in the chain. The solid line represents the sum of neutrons from SF and ($\alpha$, n) reaction, while the dashed line shows the contribution of SF only. The following decay chains are shown: 
$^{238}\rm{U}  \rightarrow \,      ^{230}\rm{Th}$ (black), 
$^{226}\rm{Ra}   \rightarrow \,      ^{206}\rm{Pb}$ (green), 
$^{235}\rm{U}   \rightarrow \,       ^{207}\rm{Pb}$ (red), 
$^{232}\rm{Th}   \rightarrow \,        ^{228}\rm{Ac}$ (blue),
$^{228}\rm{Th}   \rightarrow \,    ^{208}\rm{Pb}$ (purple).}
\label{fi:neutronyield}
\end{figure}

The radiogenic neutron production rates and energy
spectra were calculated with the SOURCES-4A software \cite{SOURCES}, with the procedure used for XENON100 \cite{xe100-nrbkg, AlexPhD}.
The neutron production rates for all the relevant materials in the background prediction are presented in table \ref{ta:neutronyield}, considering also chain disequilibrium. The neutron yield from SF ($1.1 \cdot 10^{-6}$ neutrons per decay) is included in the $^{238}\rm{U}$ column. 
For heavy nuclei, the high Coulomb barrier suppresses the ($\alpha$, n) interaction, so the neutron production is almost entirely due to SF.
The highest ($\alpha$, n) yields are from light materials, such as PTFE and the ceramic of the PMT stem. 
Among the various isotopes, we note that the neutron emission from the first part of the $^{232}$Th chain is negligible.
The energy spectra obtained for two materials, PTFE (low $Z$) and copper (high $Z$), are shown in figure \ref{fi:neutronyield}.

\begin{table}[t!]
\centering
{
\normalsize
\begin{tabular}{ | c  c  c  c  c  c |}
\hline
\textbf{Material} & \multicolumn{5}{c|}{   \textbf{Neutron yield} $[\textrm{neutrons/decay}]$ } \Tstrut\Bstrut  \\ 
\hline
 & $\mathbf{^{238}U}$ & $\mathbf{^{235}U}$ & $\mathbf{^{226}Ra}$ & $\mathbf{^{232}Th}$ & $\mathbf{^{228}Th}$ \Tstrut\Bstrut \\
\hline
\bf Stainless Steel & $1.1 \cdot 10^{-6}$ & $4.1 \cdot 10^{-7}$ & $3.1 \cdot 10^{-7}$ & $1.8 \cdot 10^{-9}$ & $2.0 \cdot10^{-6}$ \Tstrut \\
\bf PTFE (C$_2$F$_4$) & $7.4 \cdot 10^{-6}$ & $1.3 \cdot 10^{-4}$ & $5.5 \cdot 10^{-5}$ & $7.3 \cdot 10^{-7}$ & $1.0 \cdot 10^{-4}$ \\
\bf Copper & $1.1 \cdot 10^{-6}$ & $3.3 \cdot 10^{-8}$ & $2.5 \cdot 10^{-8}$ & $3.0 \cdot 10^{-11}$ & $3.6 \cdot 10^{-7}$ \\
\bf Ceramic (Al$_2$O$_3$) & $1.2 \cdot 10^{-6}$ & $1.3 \cdot 10^{-5}$ & $6.0 \cdot 10^{-6}$ & $9.2 \cdot 10^{-9}$ & $1.4 \cdot 10^{-5}$ \\
\bf Quartz (SiO$_2$) & $1.2 \cdot 10^{-6}$ & $1.9 \cdot 10^{-6}$ & $8.8 \cdot 10^{-7}$ & $6.8 \cdot 10^{-9}$ & $1.9 \cdot 10^{-6}$ \\
\bf Kovar & $1.1 \cdot 10^{-6}$ & $1.3 \cdot 10^{-7}$ & $1.2 \cdot 10^{-7}$ & $3.0 \cdot 10^{-11}$ & $1.0 \cdot 10^{-6}$ \\
\bf Cirlex (C$_{22}$H$_{10}$N$_2$O$_5$) & $1.3 \cdot 10^{-6}$ & $2.2 \cdot 10^{-6}$ & $3.5 \cdot 10^{-6}$ & $4.1 \cdot 10^{-8}$ & $2.4 \cdot 10^{-6}$ \Bstrut\\

\hline
\end{tabular}
}

\caption{Neutron production rates for the most relevant materials in the XENON1T experiment. The results are expressed as neutrons emitted for each disintegration of the parent element in the chain. To consider potential disequilibrium we separated the $^{238}\rm{U}$ and $^{232}\rm{Th}$ chains in two branches, see section \ref{se:screening} for the details.}
\label{ta:neutronyield}

\end{table}

\begin{figure}[t!]
\centering

\mbox{
\begin{minipage}[t]{0.49\textwidth}
\centering

\includegraphics[width=1.\linewidth]{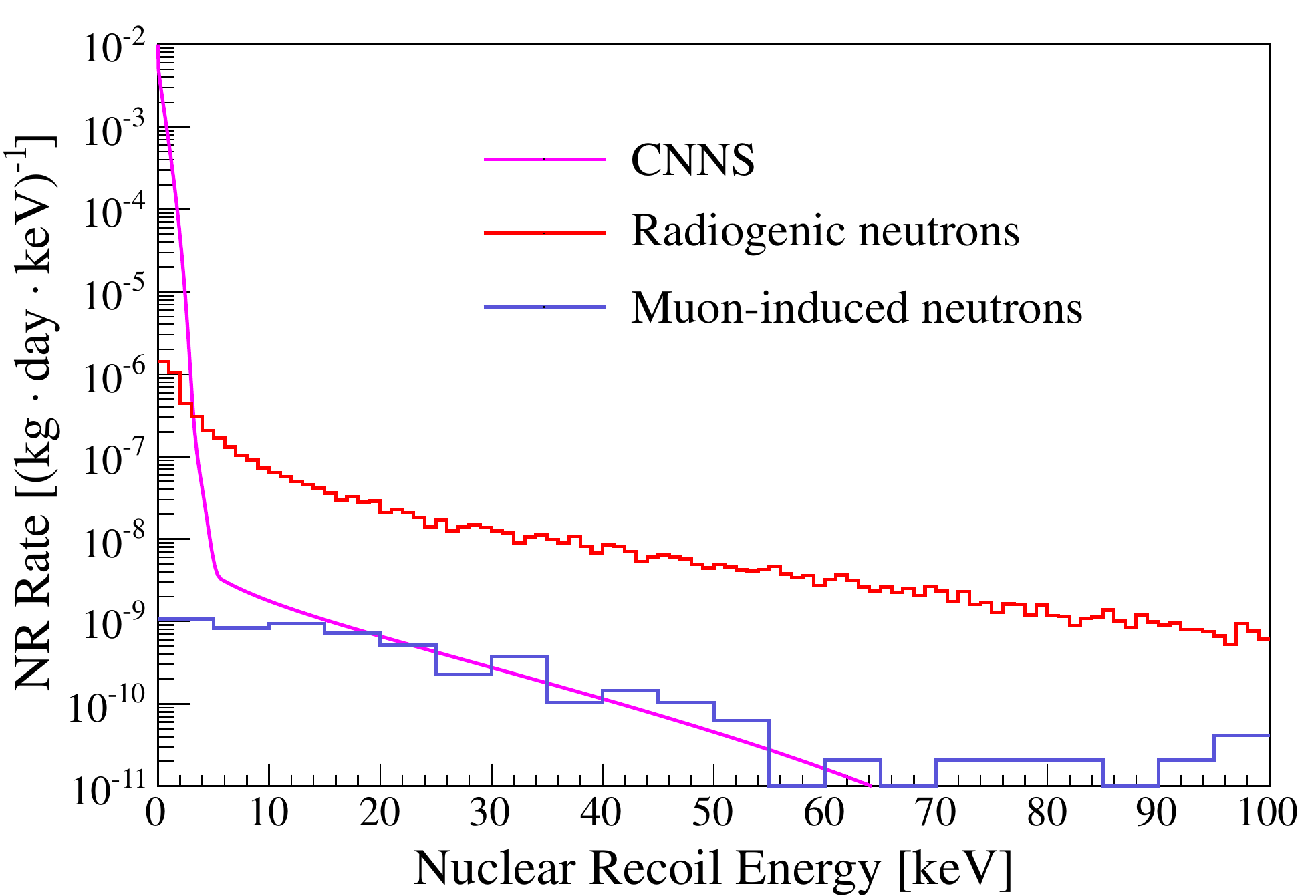}
\caption{Energy spectrum of the NR background events in $1$~t FV. 
In red we show the contribution of radiogenic neutrons from the detector components, in purple the one due to coherent neutrino-nucleus scattering, and in blue that of muon-induced neutrons.}
\label{fi:nr-spectrum}

\end{minipage}
\hspace{0.02\textwidth}
\begin{minipage}[t]{0.49\textwidth}
\centering

\includegraphics[width=1.\linewidth]{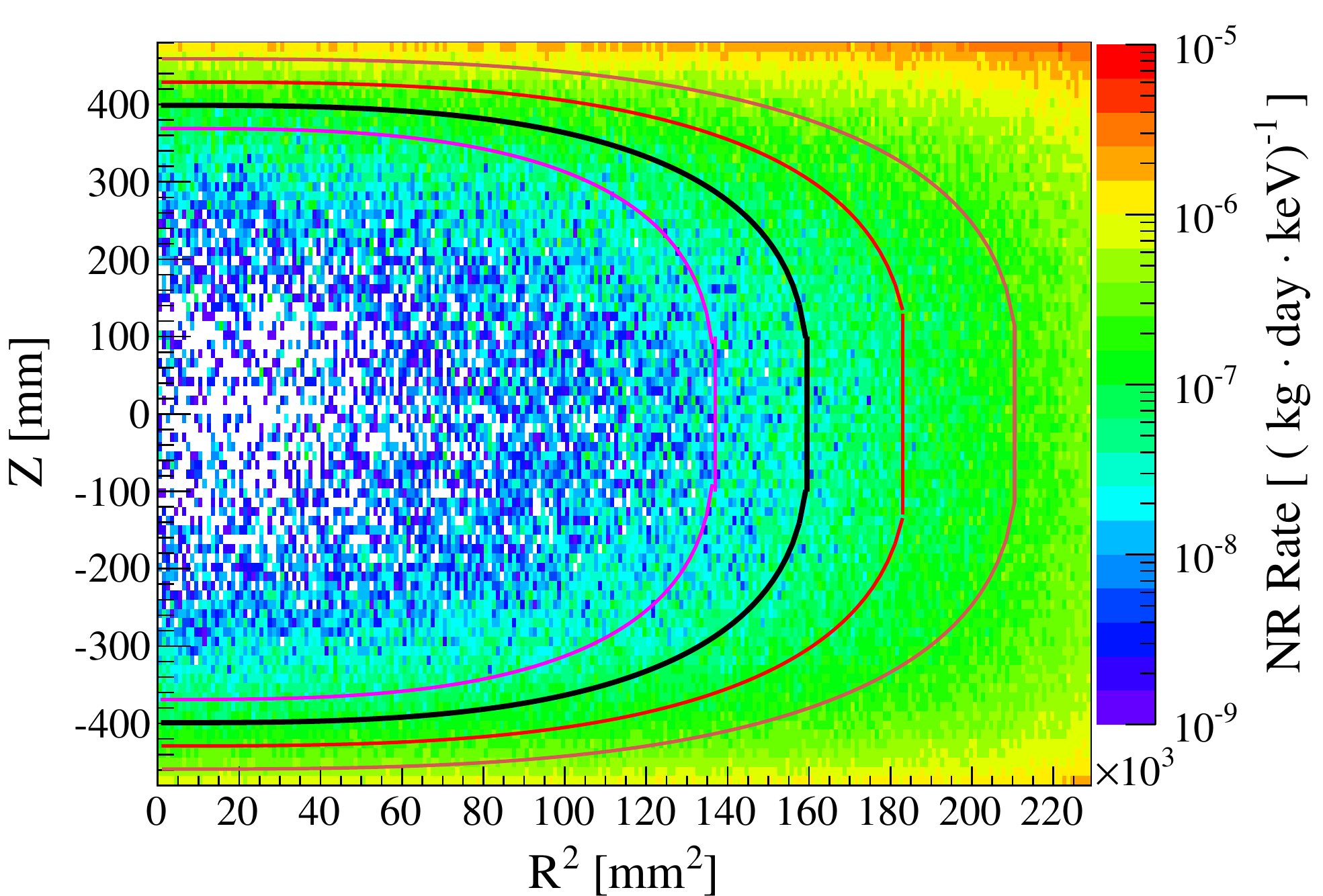}
\caption{Spatial distribution of the NR background events from radiogenic neutrons inside the active LXe volume, in the ($4$, $50$)~keV energy range. The thick black  line indicates the reference $1$~t super-ellipsoid fiducial volume. In purple, red and brown, we indicate the FV corresponding to $800$~kg, $1250$~kg and $1530$~kg, respectively. The white regions present a background rate smaller than $1 \cdot 10^{-9}$~\dru.}
\label{fi:nr-zvsr}

\end{minipage}
}

\end{figure}

We generated $10^7$ neutrons for each of the components listed in table \ref{ta:screening}, with the energy spectrum obtained with SOURCES-4A for each of the neutron sources in table \ref{ta:neutronyield}. We select the events that mimic a WIMP signal with a single elastic scatter in the whole active volume, and occur inside the fiducial volume. 
The results are scaled by the neutron yield of the material and the contamination of the component.

\begin{figure}[t!]
\centering

\mbox{
\begin{minipage}[t]{0.49\textwidth}
\centering

\includegraphics[width=1.\linewidth]{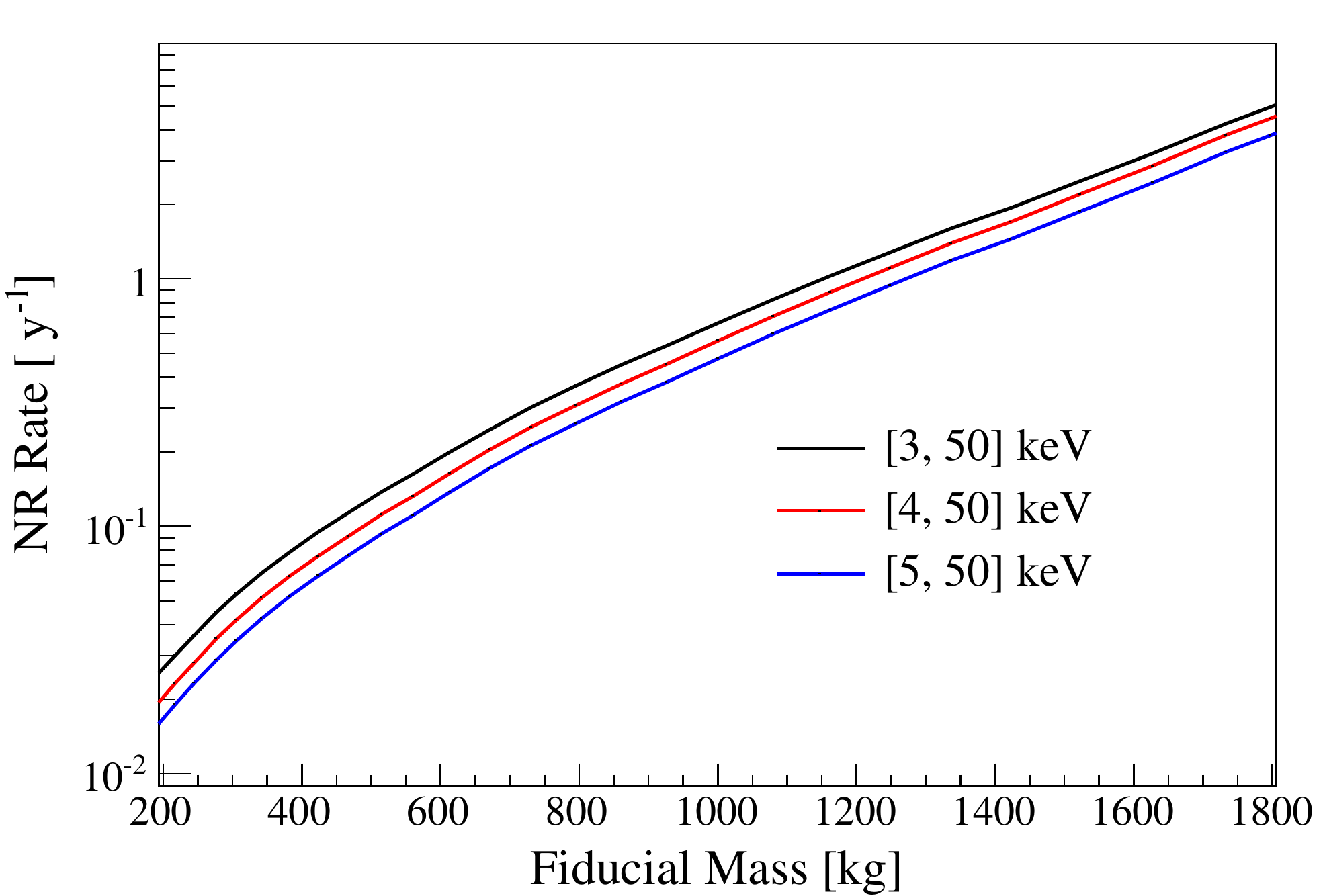}
\caption{NR background rate from radiogenic neutrons as a function of the fiducial mass, integrated in the ($3$, $50$)~keV energy range (black), ($4$, $50$)~keV (red) and ($5$, $50$)~keV (blue). The event rate is also integrated in the considered fiducial mass.}
\label{fi:nr-bkgvsmass}

\end{minipage}
\hspace{0.02\textwidth}
\begin{minipage}[t]{0.49\textwidth}
\centering

\includegraphics[width=1.\linewidth]{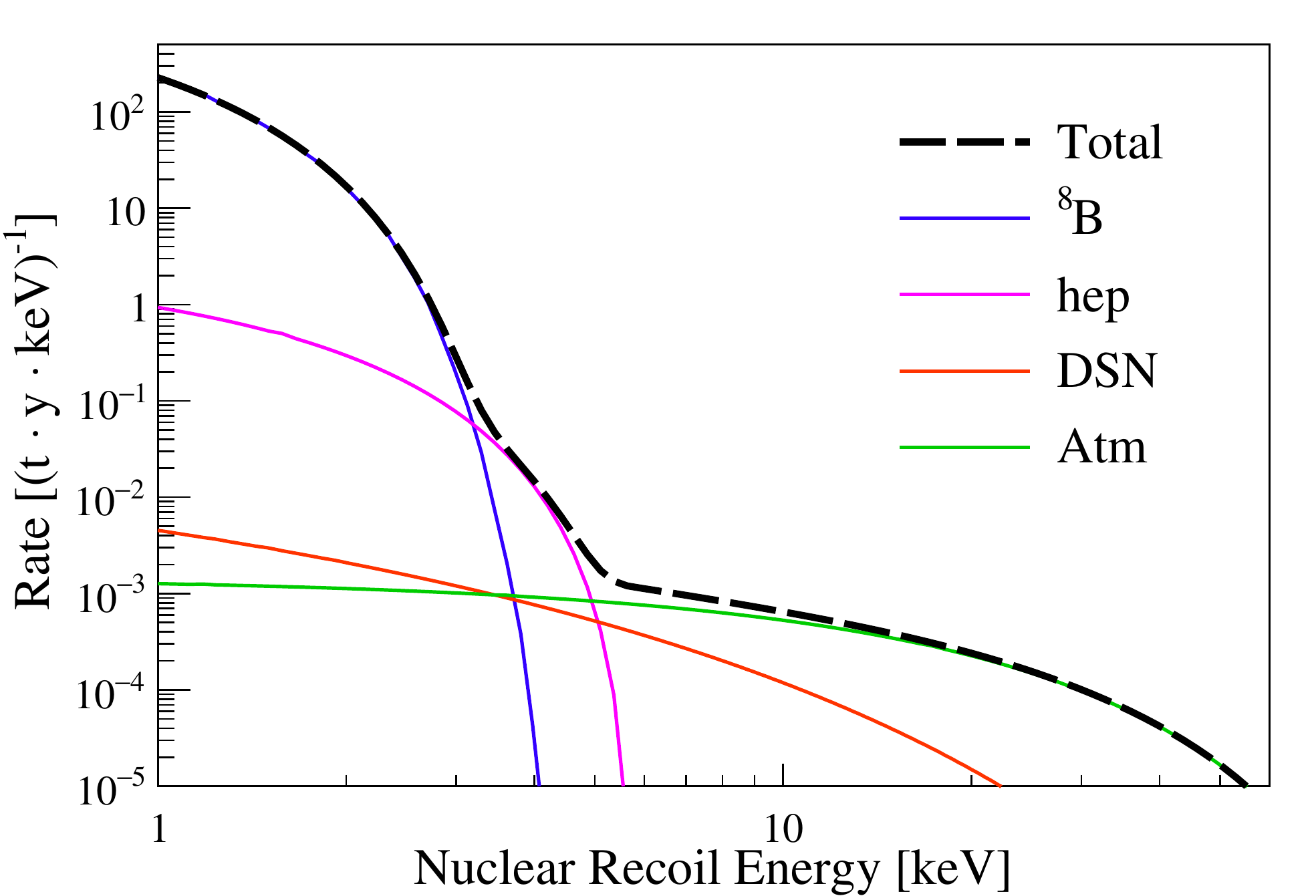}
\caption{Event rate from coherent neutrino-nucleus scattering in xenon, in the energy region of interest for WIMP search: solar ($^8$B and $hep$), diffuse supernova (DSN) and atmospheric neutrinos (Atm).}
\label{fig:CNNS}

\end{minipage}
}

\end{figure}

The energy spectrum of the total NR background from materials in a $1$~t FV is shown in figure \ref{fi:nr-spectrum}. Considering a ($4$, $50$)~keV energy range, which corresponds to the ($1$, $12$)~keV ER energy used in the ER background estimation (and contains about $70\%$ of the NRs from a $100$~GeV/$c^2$ WIMP), the background rate is $(0.6 \pm 0.1)$ y$^{-1}$ in $1$~t. The uncertainty is dominated by the $17\%$ systematics in the neutron yield from the SOURCES-4A code \cite{SOURCES}. The largest fraction of background events comes from the cryostat SS ($28\%$), followed by the PMTs ($26\%$, mostly from the ceramic stem), PTFE ($20\%$), SS of the TPC ($10\%$) and of the reservoir ($7\%$), PMT bases ($5\%$).
A cross-check of the prediction of the NR background from radiogenic neutrons was performed \cite{LRT2013} with an independent code \cite{ZhangMei}. The results were found in agreement within the assumed systematic uncertainty, with SOURCES-4A predicting the largest background.
Neglecting the background from the materials where only upper limits were found, the total event rate from neutrons decreases by about $20\%$. However, for this sensitivity study, we assumed the most conservative values obtained with SOURCES-4A, including the upper limits. 
The spatial distribution of the NR background events inside the active volume, in the ($4$, $50$)~keV energy region is shown in figure \ref{fi:nr-zvsr}. In figure \ref{fi:nr-bkgvsmass} we show the background rate as a function of the fiducial mass, for three values of the lower edge of the energy region of interest: $3$, $4$ and $5$~keV. The variation with respect to the central one is $\sim 20\%$.




\subsection{Muon-induced neutrons}
Neutrons are also produced by the interaction of cosmic muons with the rock and concrete around the underground laboratory, and with the detector materials. The neutron energy extends up to the GeV range, so they can penetrate even through large shields and reach the sensitive part of the detector, mimicking a WIMP interaction.
To protect XENON1T from this background, the detector is placed inside a cylindrical water tank, $9.6$~m in diameter and $\sim 10$~m in height, which acts as a shield against both external neutrons and $\gamma$-rays. In addition, the tank is instrumented with 84 8-inch diameter PMTs, Hamamatsu R5912ASSY, to tag the muon and its induced showers through the detection of the Cherenkov light produced in water. 
The details of the MC simulation to model the production, propagation and interaction of the muon-induced neutrons, and the performance of the muon veto are described in \cite{XENON1T-MuonVeto}. We are able to tag $> 99.5\%$ of the events where the muon crosses the water tank and $> 70\%$ of those where the muon is outside the tank, but the neutrons enter together with their associated showers. Assuming conservative values for the muon-induced neutron yield \cite{MeiHime}, and considering the effect of both the passive shielding of the water and the active veto, the surviving neutron background is $< 0.01$ y$^{-1}$ in $1$~t FV.
 This is negligible and thus is not considered in the estimation of the sensitivity. The NR energy spectrum produced by muon-induced neutrons is shown in figure \ref{fi:nr-spectrum}.

\subsection{Coherent neutrino-nucleus scattering}
Neutrinos contribute to the NR background through CNNS.
We followed the approach developed in \cite{Billard} to calculate the expected rate of events considering solar neutrinos (from all the various reactions in the Sun), diffuse supernova and atmospheric neutrinos. The dominant contribution in the energy region of interest for the dark matter search comes from the $^{8}\rm{B}$ and $hep$ neutrinos from the Sun, while the event rate from higher energy neutrinos (diffuse supernovae and atmospheric) is orders of magnitude smaller, as shown in figure \ref{fig:CNNS}. The integral event rate above an energy threshold of $3$, $4$ and $5$~keV is very small: $9.1 \cdot 10^{-2}$~\tyu$\,$, $1.8 \cdot 10^{-2}$~\tyu$\,$  and $1.2 \cdot 10^{-2}$~\tyu, respectively. However, given the very steep energy spectrum of NR events from CNNS, it is also necessary to estimate the event rate starting from a lower energy threshold. For instance, the event rate above $1$~keV is $\sim 90$~\tyu, and due to the small number of detected photons, the poissonian fluctuations in the generated signal can allow detection of the low energy events, as described in section \ref{S:Conv}. 
The uncertainty for CNNS events is $14\%$, coming mainly from the uncertainty of the $^{8}\rm{B}$ neutrino flux from the Sun \cite{Serenelli:2011py}.

\subsection{Summary of NR backgrounds}
The NR background spectrum in $1$~t FV is summarized in figure \ref{fi:nr-spectrum}: the different contributions have been evaluated in the NR energy region ($4$, $50$)~keV, which corresponds to the one used for ERs when taking into account the different response of LXe to ERs and NRs.  
The main contribution is due to radiogenic neutrons which produce $(0.6 \pm 0.1)$~\tyu, calculated using the contaminations of the detector materials.
The second one comes from neutrinos through their coherent scattering off xenon nuclei: their rate in the same energy region is very small,  $(1.8 \pm 0.3) \cdot 10^{-2}$~\tyu, but will become relevant once the detector response and the energy resolution are taken into account (in section \ref{S:Conv}).
Finally, the background from muon-induced neutrons is reduced to less than 0.01~\tyu $\,$ by the water Cherenkov muon veto, and its contribution is neglected in the study of the experiment sensitivity.

\section{Light yield}
\label{S:LCE}

The main goal of a dual-phase xenon TPC such as XENON1T is to detect low intensity VUV light signals, produced either directly (S1) or through proportional scintillation (S2), in order to be sensitive to low energy NRs.
The light collection efficiency (LCE) is defined as the fraction of emitted photons reaching the PMTs. It depends on the position of the interaction in the active volume, and on the optical properties of LXe and of the materials around it. 

Several measurements of the optical properties of LXe have been performed throughout the years, in the following for all of them we refer to the emission wavelength of $178$~nm.
For the refractive index, values were measured between $1.565 \pm 0.002 \pm 0.008$ \cite{Barkov1996} and $1.69 \pm 0.02$ 
\cite{Solovov2004}: in the simulation, we use the average value of $1.63$. 
The second optical parameter required is the Rayleigh scattering length, which affects the mean free path of photons. 
We adopt in the simulation the theoretical value of  $30$~cm \cite{Seidel}, which is in agreement with the lowest measured values, ranging from $\sim30$~cm to $\sim50$~cm \cite{Braem1992,Chepel1994,Solovov2004}.
The last parameter of interest is the absorption length which mostly depends on the amount of impurities present in the LXe ($\mathrm{O_2}$ and, more importantly, $\mathrm{H_2O}$) and is therefore dependent on the performance of the purification system.
In XENON1T we adopt a similar system to the one used in XENON100 \cite{xe100-instrument} and we aim to 
reach a sub-ppb level for those impurities. 
Knowing that in \cite{Baldini2005} an absorption length of $1$~m was achieved with a $100$~ppb level of water, and assuming an inversely proportional scaling, we consider a conservative value of $50$~m.

Another material whose optical properties must be properly defined is the quartz of the PMT window, as it governs the amount of light transmitted to the photocathode and, thus, the LCE.
Using measurements  from \cite{Malitson1965} and \cite{Kitamura2007}, we chose a refractive index of $1.59$ for a wavelength of $178$~nm. The photocathode is modeled as a fully-opaque thin layer, placed on the inner side of the PMT window where photons are absorbed.
Apart from the cut-outs needed to host the PMTs, the inner surface of the TPC is entirely made of PTFE in order to reflect as much VUV light as possible. The reflectivity of the PTFE strongly depends on the surface treatment, thus we studied the LCE with PTFE reflectivity values of $90\%$, $95\%$ and $99\%$. A very good reflectivity  ($> 99\%$) has been obtained in the LUX detector \cite{LUX-tech}. 
The layer of GXe located between the LXe and the top PMT array is characterized by a refractive index equal to 1.

Finally, we also modeled the five electrodes used to define the electric field within the TPC.
Four of them are hexagonally etched meshes: the top and bottom screening meshes have an optical transparency of $94.5\%$, while that of the anode and gate meshes is $93\%$. 
The cathode is made of thin wires and offers an optical transparency of $96\%$.
In our simulation framework, the electrodes are modeled as a $200$~$\mathrm{\mu m}$ thick disks with absorption length calculated to match their respective transparency at normal angle. Furthermore, the refractive index of each electrode is the same as its surrounding material (LXe or GXe): the reflectivity is, therefore, conservatively neglected here. 

\begin{figure}[t]
\centering

\mbox{
\begin{minipage}[t]{0.49\textwidth}
\centering

\includegraphics[width=1.\linewidth]{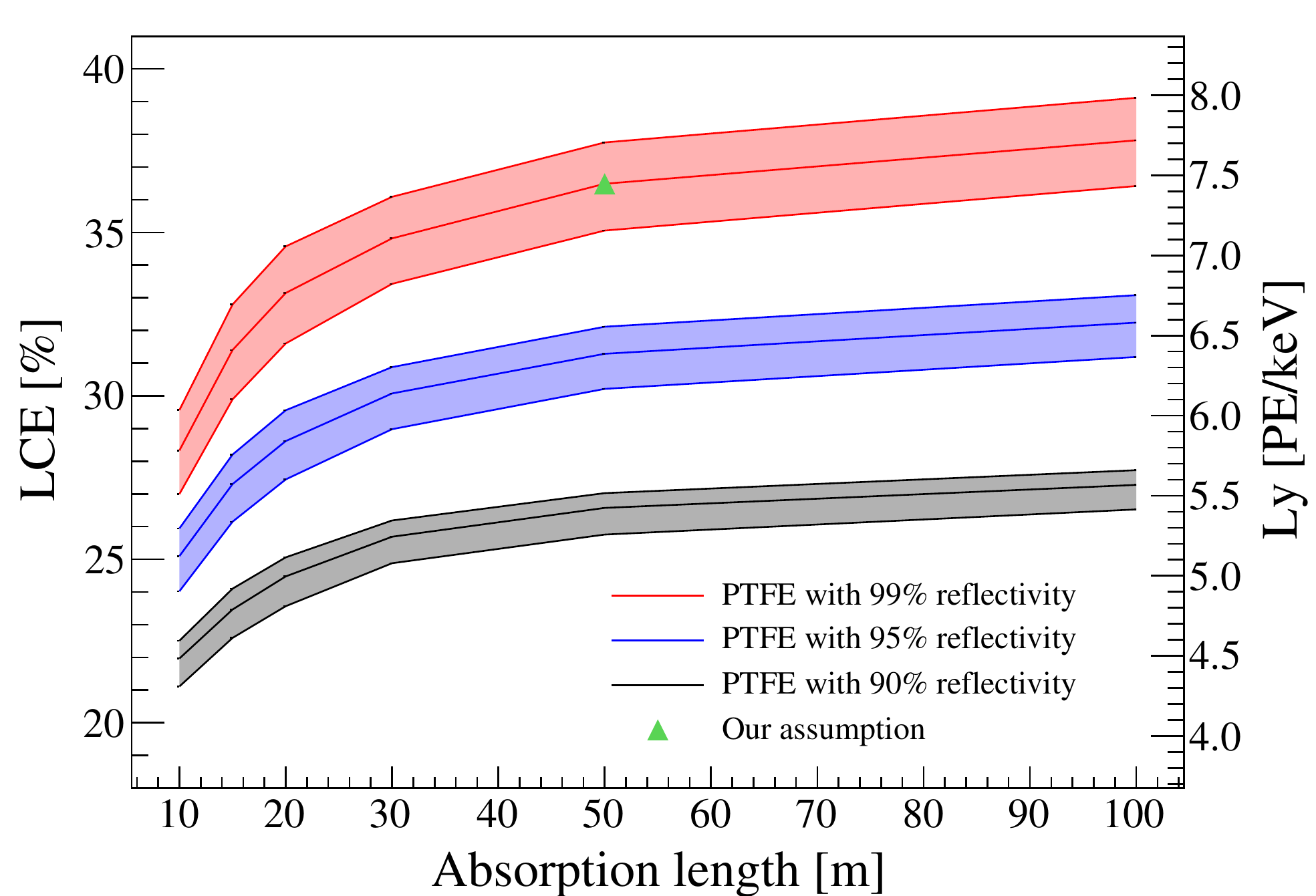}
\caption{Average light collection efficiency in the TPC as a function of the absorption length of LXe, for different PTFE reflectivities. For each PTFE reflectivity, the upper line corresponds to a LXe refractive index of $1.565$, while the lower line is for $1.69$.  In the right side of the Y axis we show the corresponding light yield. The green triangle marks the values of the optical parameters assumed for the sensitivity study presented in the following sections. }
\label{fi:lceVariations}

\end{minipage}
\hspace{0.02\textwidth}
\begin{minipage}[t]{0.49\textwidth}
\centering

\includegraphics[width=1.\linewidth]{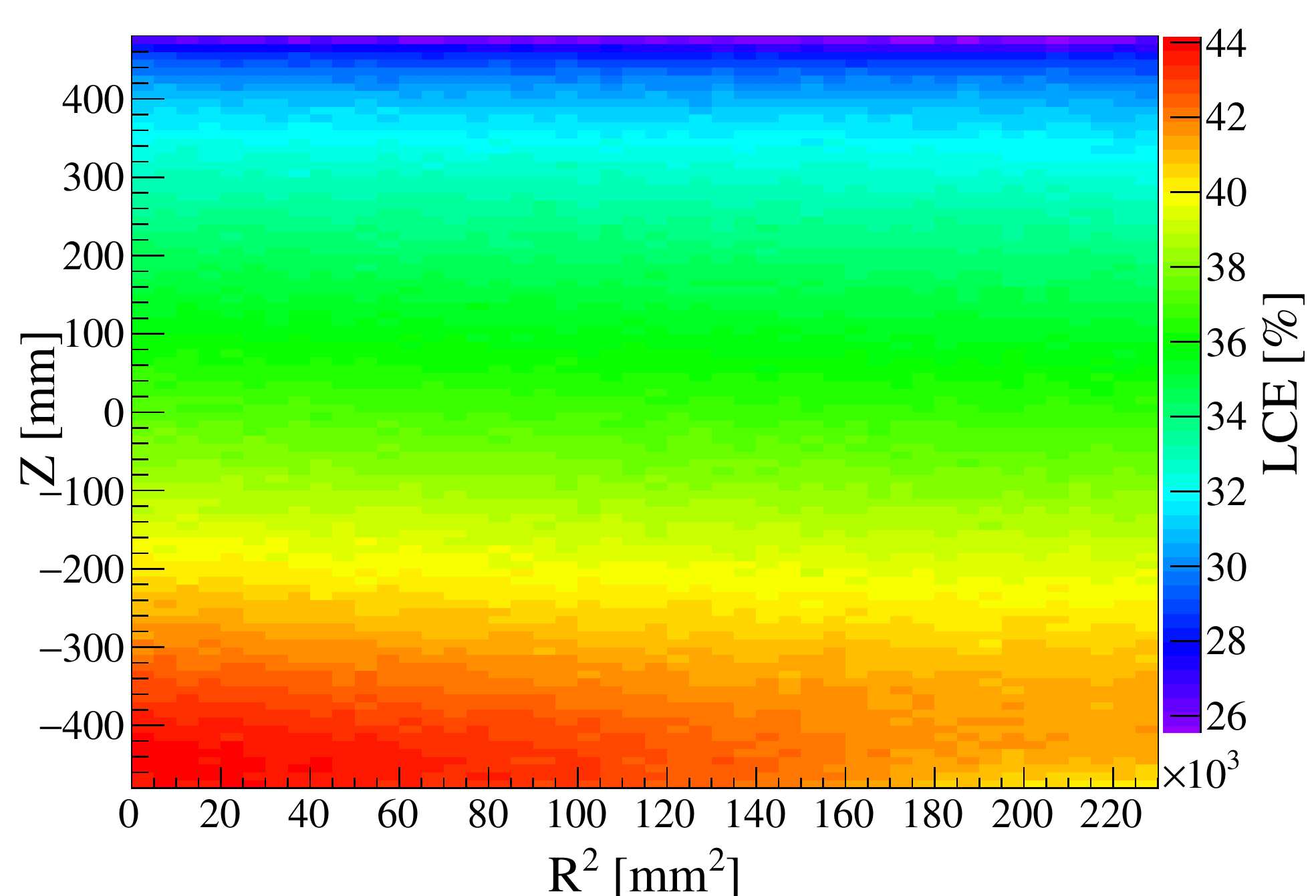}
\caption{LCE as a function of the position in the TPC, calculated with the parameters assumed for the XENON1T model (green triangle in figure \ref{fi:lceVariations}). 
The LCE is the fraction of generated photons reaching a PMT photocathode.
}
\label{fi:lceMap}

\end{minipage}
}

\end{figure}

In this study, photons are generated uniformly and isotropically in the full volume of the TPC, and the LCE is  calculated for each individual $\mathrm{(R^{2},Z)}$ pixel, using the axial symmetry of the TPC.
In figure \ref{fi:lceVariations}, the LCE averaged over the active LXe volume is shown for different PTFE reflectivities and different absorption lengths: improving the reflectivity from $90\%$  to $99\%$ leads to a $37\%$ gain in LCE, for a $50$~m absorption length.
Furthermore, increasing the absorption length from $10$~m to $50$~m raises the LCE by $28\%$, for a fixed PTFE reflectivity of $99\%$. 
The variation of LCE inside the TPC is shown in figure \ref{fi:lceMap}.
Given the internal reflection occurring at the liquid/gas interface, the LCE is higher close to the bottom PMT array, in particular in the center of the TPC, and decreases when moving closer to the anode due to the increase in path length, leading to absorption of photons.

The light yield $\mathcal{L}_y(\vec{r})$ is defined as the specific number of detected photoelectrons (PE) per keV, and it is traditionally referred to the $122$~keV $\gamma$ emitted by a $^{57}\rm{Co}$ source, at zero electric field. The average photon yield ${Ph_y}$ of this $\gamma$ line is estimated with a phenomenological model in NEST \cite{NEST2011, NEST2013} as $63.4$~ph/keV.
$\mathcal{L}_y(\vec{r})$ is calculated as:
\begin{equation}\label{eq:LyDef}
\mathcal{L}_y(\vec{r}) = f_{PE}(\vec{r}) \cdot  {Ph_y} = \text{LCE}(\vec{r}) \cdot {QE} \cdot {CE} \cdot  {Ph_y}
\end{equation}
where $f_{PE}$ is the probability for an emitted photon to produce a photoelectron, LCE is the light collection efficiency, $QE$ is the average quantum efficiency of the PMTs ($35\%$) \footnote{The quoted $QE$ has been measured at room temperature. However in \cite{Lyashenko:2014rda} it was reported an increase of $QE$ at the LXe temperature, so the performances can be even better. In this study we conservatively assume the room temperature $QE$.}, and $CE$ is the average collection efficiency from the photocathode to the first dynode ($90\%$) \cite{PMTperformances1}.
The light yield is shown in the right vertical axis of figure \ref{fi:lceVariations}.

In the following sections, we will estimate the sensitivity of XENON1T assuming a $99\%$ PTFE reflectivity, $50$~m in absorption length and the central value of the LXe refraction index, as shown by the green triangle in figure \ref{fi:lceVariations}: with this choice we obtain an average light yield in the TPC of $7.7$~PE/keV at zero field. This result is about twice the one obtained in XENON100 ($3.8$~PE/keV \cite{xe100-run10}) and close to the one measured in LUX ($8.8$~PE/keV \cite{LUX-results}).


\section{Conversion of energy deposition into light and charge signals}
\label{S:Conv}

To convert the energy deposited in the active LXe ($E_d$) into the light (S1) and charge (S2) signals, we first need a model which predicts the amount of generated photons and electrons. Then all the efficiencies in collecting the signals must be taken into account, together with the signal statistical fluctuations.   
Even if the design goal is to use an electric field in the TPC of $1$~kV/cm, in the following we will conservatively assume $530$~V/cm, as in XENON100, so that we can apply directly some of the measurements obtained with that detector, in particular the specific charge yield for NR $\mathcal{Q}_\mathrm{y}$.

\subsection{Generation of photons and electrons}
ER and NR present different scintillation and ionization yields, hence they are treated in the simulation using two separate models.

For ERs, we used the approach developed in NEST (version 0.98) \cite{NEST2011, NEST2013}: first the total number of quanta $\mean{N_Q} = (73 \cdot E_d/\mathrm{keV})$ is calculated and smeared using a Gaussian distribution with a 0.03 Fano factor \cite{Seguinot:1995uf}.
Then they are shared between $N_{ph}$ photons and $N_{el}$  electrons, such that  $N_Q =  N_{ph} + N_{el}$, to ensure the anti-correlation between the two signals. The amount of photons and electrons is evaluated following the Doke-Birks \cite{Doke:1988dp} recombination model or the Thomas-Imel \cite{Thomas:1987zz} model for short length tracks, corresponding to energies smaller than $\sim (10-15)$~keV. The recombination is calculated directly in the GEANT4 simulation, on a event-by-event basis, considering the $dE/dX$ of each particle for Doke-Birks, and its energy for the Thomas-Imel model. The fluctuations on $N_{ph}$ (and the corresponding ones for $N_{el}$) are properly taken into account using sampling from a Binomial distribution. 

Traditionally, for NR, the photon yield is parameterized in terms of the so-called relative scintillation efficiency in LXe ($\mathcal{L}_\mathrm{eff}$). Similarly to the light yield described in equation \ref{eq:LyDef}, also $\mathcal{L}_\mathrm{eff}$ is defined with reference to $Ph_y$, the light emitted by the $122$~keV $\gamma$ from $^{57}\rm{Co}$.
We used a strategy similar to that described in \cite{MarcW}, obtaining the average number of photons $\mean{N_{ph}^{NR}}$ as:

\begin{equation}
\nonumber
\mean{N_{ph}^{NR}}  = E_d \cdot  \mathcal{L}_\mathrm{eff} \cdot  {Ph_y} \cdot  \mathrm{S_{NR}} 
\end{equation}
where $\mathrm{S_{NR}} = 0.95$ is the light yield suppression factor for NR, due to the electric field. As in the previous analyses of the XENON100 experiment, $\mathcal{L}_\mathrm{eff}$ is taken from an average of the various direct measurements, as shown in figure 1 in \cite{xe100-run8}.
The average number of electrons is obtained from the charge yield ($\mathcal{Q}_\mathrm{y}$) measured in \cite{MarcW} as
 \begin{equation}
\nonumber
 \mean{N_{el}^{NR}}  = E_d \cdot  \mathcal{Q}_\mathrm{y}
\end{equation}
In NRs, a large fraction of the deposited energy is dissipated into heat and is not available to generate light and charge signals, therefore $\mean{N_Q^{NR}} =  \mean{N_{ph}^{NR}} + \mean{N_{el}^{NR}}$ is smaller than the $\mean{N_Q}$ obtained for ER of the same energy. We model the fluctuations in the available quanta with a sampling from a Binomial distribution with probability $f_{NR} =  \mean{N_Q^{NR}} / \mean{N_Q}$:  
 \begin{equation}
\nonumber
N_{Q}^{NR} = \mathrm{Binomial} (N_Q,~  f_{NR}).
\end{equation}
The number of photons $N_{ph}$ is obtained from $N_{Q}^{NR}$  with another Binomial sampling with probability $f_{ph} =  \mean{N_{ph}^{NR}} / \mean{N_Q^{NR}}$, 
 \begin{equation}
\nonumber
N_{ph} = \mathrm{Binomial}(N_Q^{NR},~  f_{ph}).
\end{equation}
The number of electrons is given by $N_{el} = N_Q^{NR} - N_{ph}$. Hence we also consider the anti-correlation between the light and charge signal for NRs.
  
\subsection{Generation of S1 and S2 signals}
\label{S:S1andS2}

\begin{figure}[t!]
\centering\includegraphics[width=0.49\linewidth]{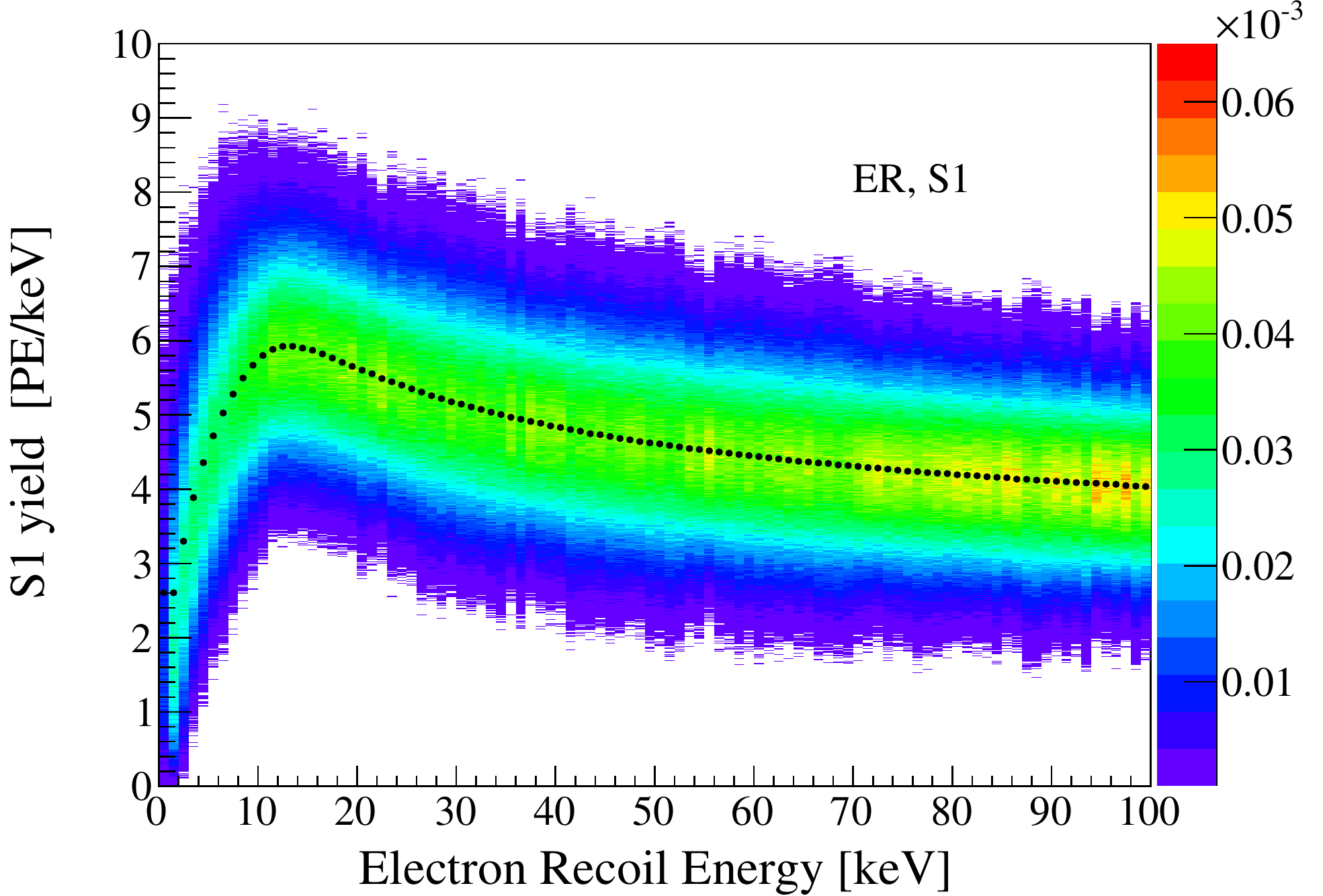}
\centering\includegraphics[width=0.49\linewidth]{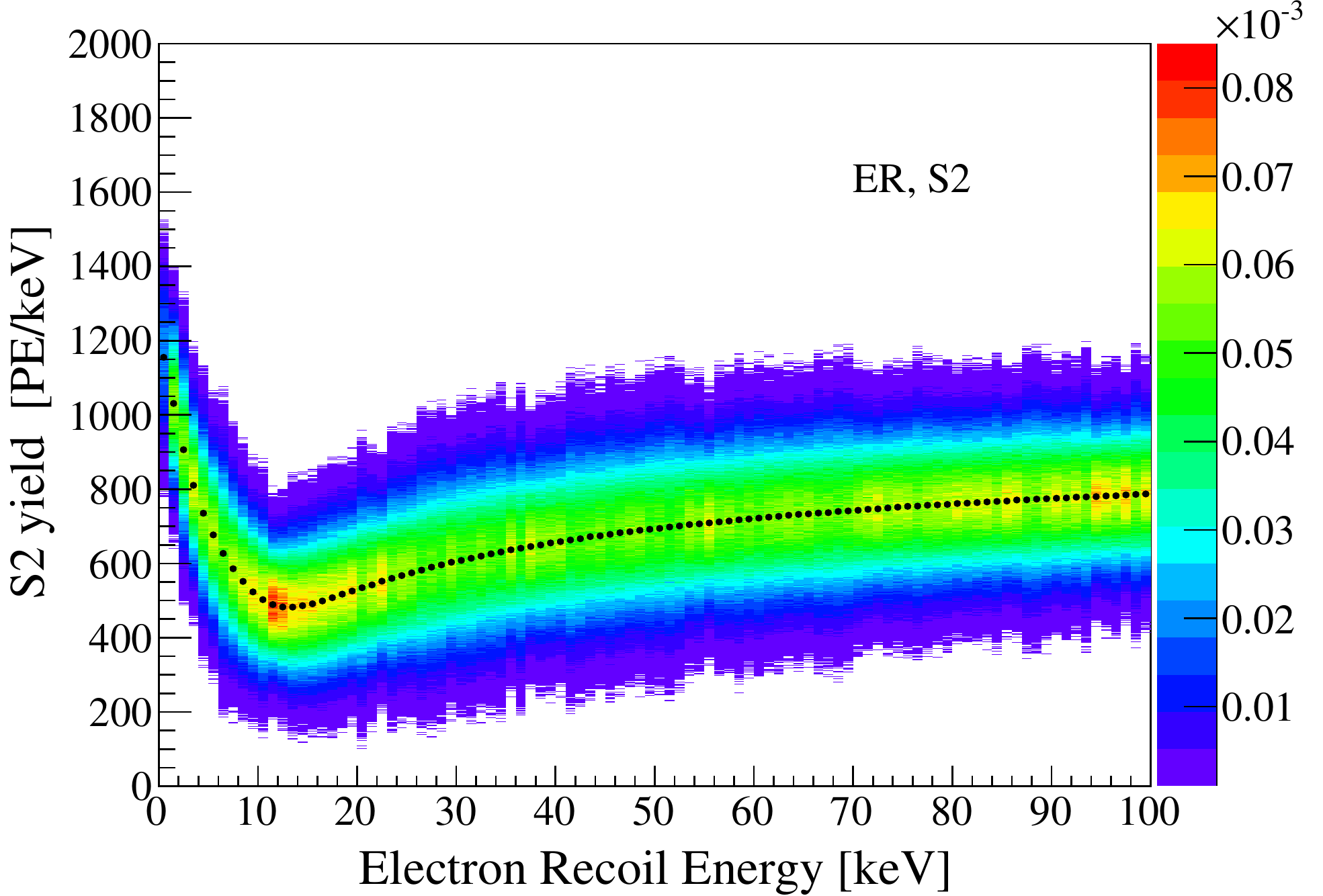}
\centering\includegraphics[width=0.49\linewidth]{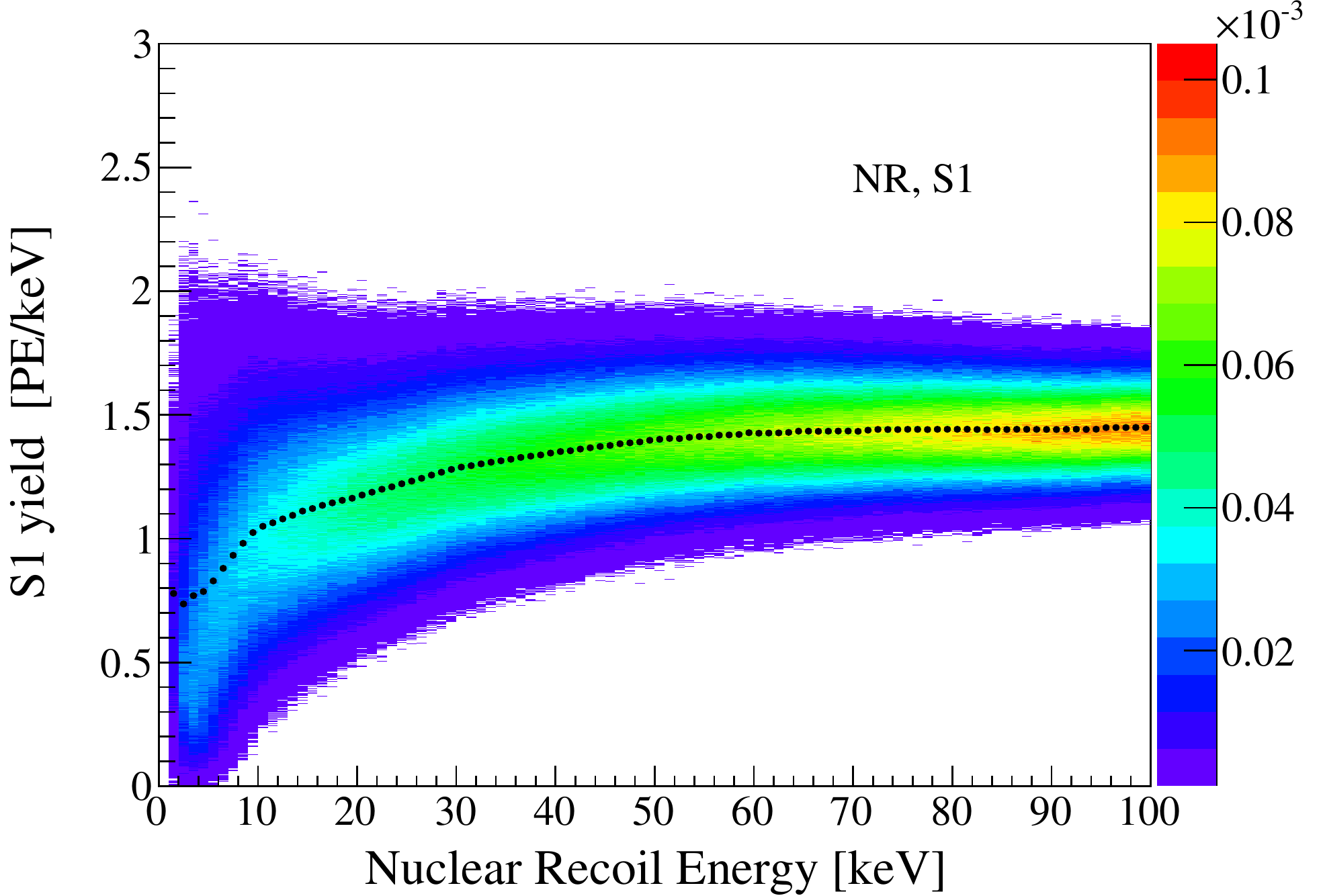}
\centering\includegraphics[width=0.49\linewidth]{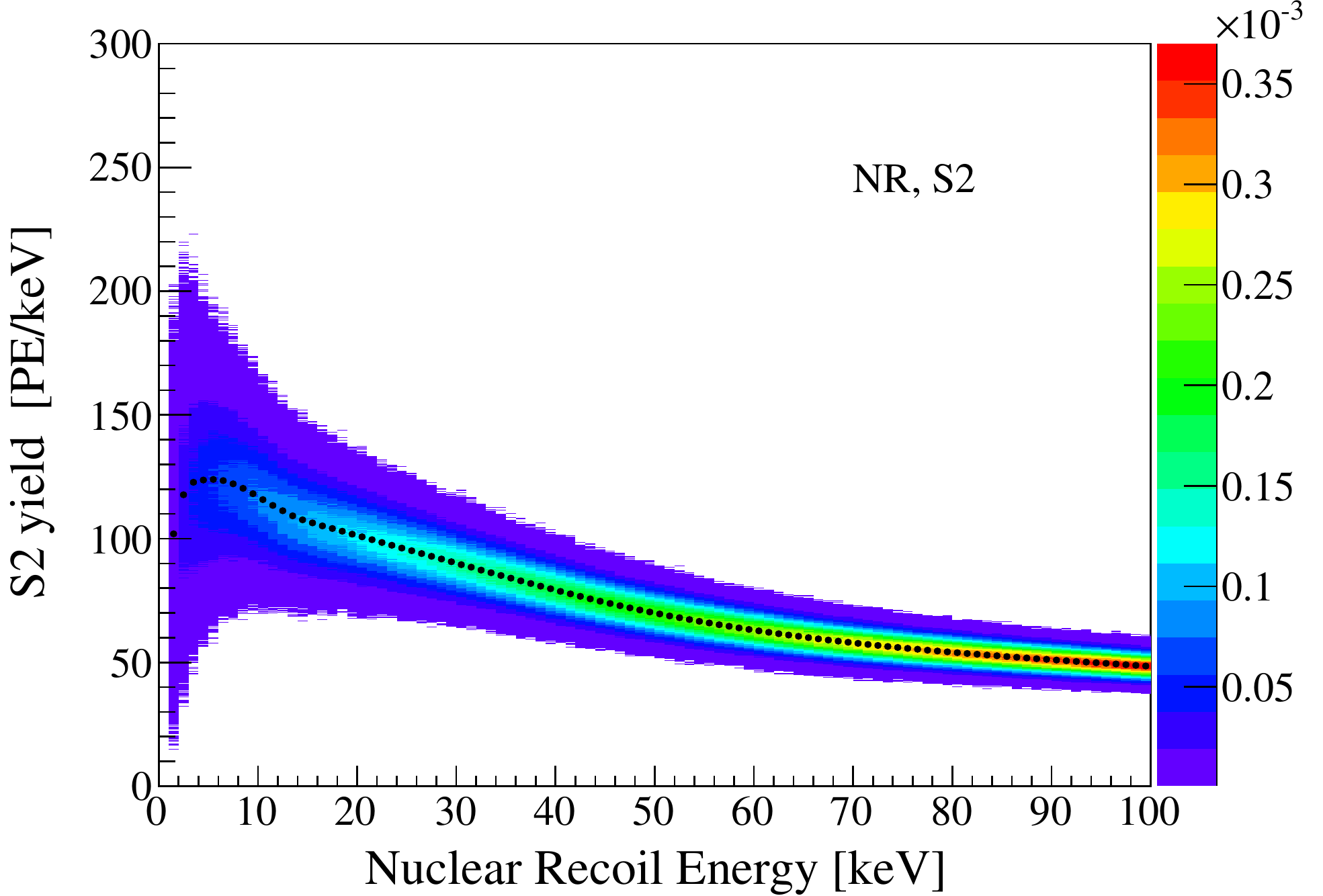}
\caption{Distribution of the S1 (left) and S2 (right) yield as a function of the deposited energy: top row is for ERs, bottom row for NRs. The z-axis for all the plots is expressed by the colors in terms of normalized units, the same number of events has been generated in each 1 keV slice of the x-axis. The black dots represent the average value of the signal yield in each energy slice.}
\label{fi:s1s2yield}
\end{figure}

Photons are converted into the S1 signal by applying the position-dependent light collection efficiency derived in section \ref{S:LCE}, see eq \ref{eq:LyDef}.
The average value of $f_{PE}$, over the whole TPC, is $\sim 12\%$, corresponding to an S1-yield (defined as the number of detected PE normalized by the deposited energy) of $4.6$~PE/keV for a $122$~keV $\gamma$ at $530$~V/cm. 
The fluctuations are taken into account considering a Binomial sampling from the number of generated photons to the detected PE: 
 \begin{equation}
\nonumber
N_{PE} = \mathrm{Binomial}(N_{ph}, ~ f_{PE}(\vec{r})).
 \end{equation}
To reproduce the response of the PMT, we apply to each single PE a convolution with a Gaussian distribution with width $0.4$ \cite{PMTscreening}: 
 \begin{equation}
\nonumber
\mathrm{S1}=\mathrm{Gauss}(N_{PE}, ~0.4\cdot \sqrt{N_{PE}}).
 \end{equation}
Then the result is corrected for the average value of the light collection efficiency in that position.
The distribution of the S1-yield as a function of $E_d$ is shown in figure \ref{fi:s1s2yield} (left column).

For the charge signal, the attenuation of electrons as they drift towards the anode is accounted for by considering the drift time $t_d(z)$ corresponding to the interaction position, and the level of impurities in the TPC, parameterized by the electron lifetime $\tau_e$: 
\begin{equation}
\nonumber
 N'_e = \mathrm{Binomial}(N_e, f_e), ~ ~ ~  \mathrm{with }~ f_e = \exp^{-(t_d(z) / \tau_e)}. 
 \end{equation}
In this study, however, we considered no electron absorption, assuming the performance of the detector after a complete purification from electronegative impurities.
 Then the S2 signal is generated assuming a full extraction efficiency (as obtained in XENON100), a mean amplification of 
 $20$~PE/e$^-$, and a Gaussian smearing with width $7$ ~PE/e$^-$  \cite{xenon100-singleelectrons}:
  \begin{equation}
\nonumber 
 \mathrm{S2} = \mathrm{Gauss}(20\cdot N'_e, ~7 \cdot \sqrt{N'_e}).
\end{equation}
The energy dependency of the S2-yield is shown in figure \ref{fi:s1s2yield} (right column).

The description of the full simulation of the digitization of the S1 and S2 signals, to reproduce their shape in the waveform, so that they can be analyzed with the same software chain as the real data from the detector, is beyond the scope of this work.

\subsection{Signal and background distributions in S1}

\begin{figure}[t!]
\centering\includegraphics[width=0.9\linewidth]{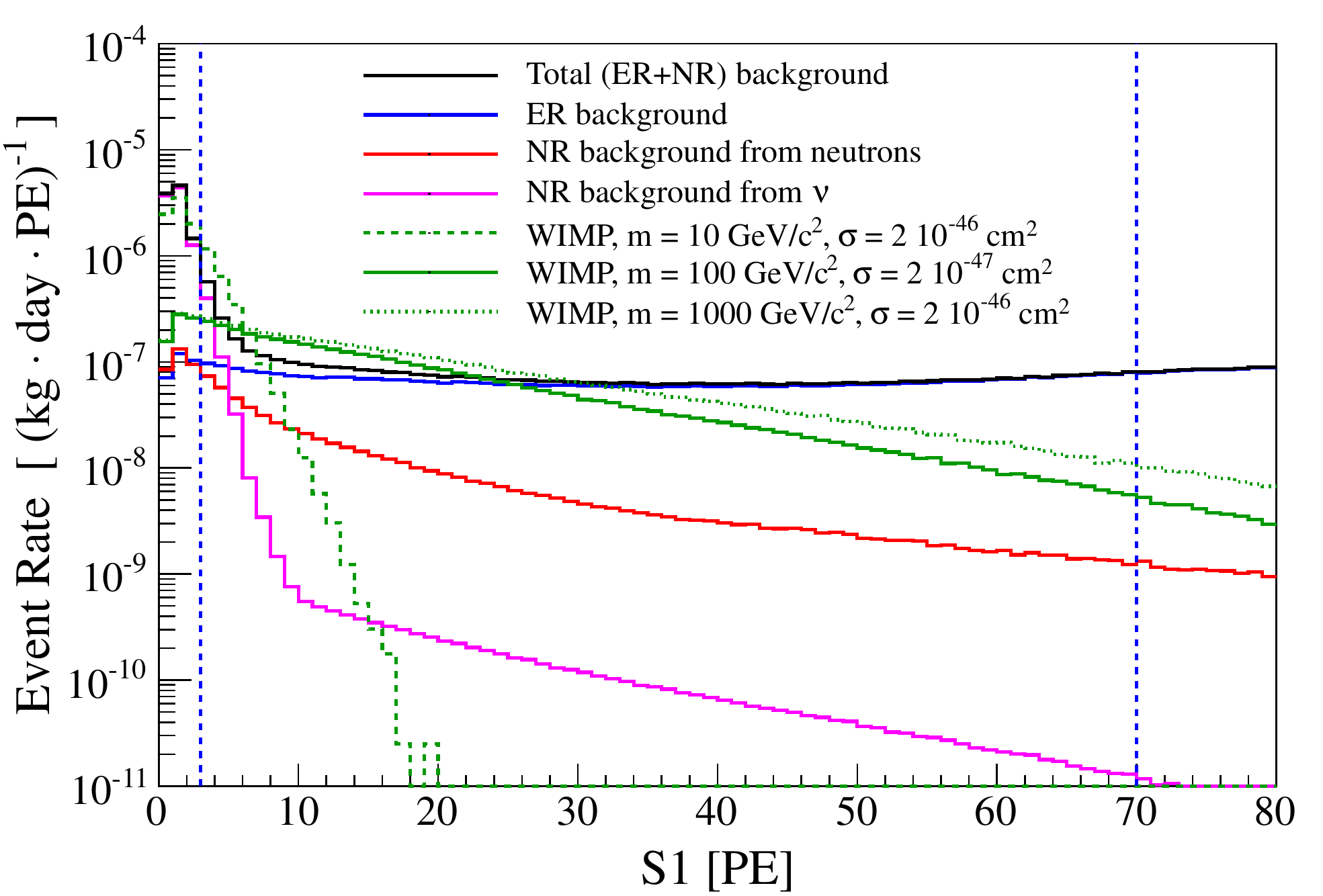}
\caption{Spectrum of the total background as a function of S1 (black) and of its components: ERs (blue), NRs from radiogenic neutrons (red) and NRs from CNNS (purple). NR spectra for three examples of WIMP signals (green):  mass $m_{\chi}=10$~GeV/$c^2$ and cross section $\sigma = 2 \cdot 10^{-46}$~cm$^2$ (dashed),  $m_{\chi}=100$~GeV/$c^2$ and $\sigma = 2 \cdot 10^{-47}$~cm$^2$ (solid), $m_{\chi}=1000$~GeV/$c^2$ and $\sigma = 2 \cdot 10^{-46}$~cm$^2$ (dotted). The vertical dashed blue lines delimit the S1 region used in the sensitivity calculation. In this plot we select the events with S2 $> 150$ PE, and assume a $99.75\%$ ER rejection with a flat $40\%$ NR acceptance.  }
\label{fi:finalbkg-s1}
\end{figure}

After the conversion of the deposited energy into the detector signals, the ER and NR backgrounds, as well as the signal from WIMPs, can be shown together in the same energy scale. In this study, we estimate the NR energy using information only from S1, as it was done in previous XENON100 analyses \cite{xe100-analysis}. In figure \ref{fi:finalbkg-s1}, the total background in $1$~t FV is presented as a function of S1, together with the separate contributions of ERs and NRs. We require to have an S1-S2 pair, with S2 $> 150$~PE (the same XENON100 trigger threshold), even if with the new XENON1T DAQ we expect an improved S2 threshold. We assume an ER rejection efficiency of $99.75\%$ with a flat $40\%$ NR acceptance.
The total background is dominated by ERs, except below $\sim 5$~PE, where the NRs from CNNS become more relevant.
For comparison, we superimpose also the NR spectrum
induced by WIMPs for three different masses and cross-sections, with the halo properties assumed in \cite{xe100-run10}.

\section{Sensitivity prediction for XENON1T}
\label{S:Sens}

We now calculate the sensitivity of the XENON1T experiment based on the background prediction and the conversion from deposited energy into observable signals presented in the previous sections.
The sensitivity is defined as the median upper limit we would obtain in repeated experiments with only background present and null signal. 
For statistical inference, in particular the calculation of upper limits, we use the Profile Likelihood Ratio method \cite{CCGV}. 
The approach was already used for the analysis of the XENON100 data \cite{xe100-run8, xe100-run10} and is described in detail in \cite{xe100-pl, xe100-analysis}.

The observables for the analysis are the prompt scintillation signals $S1$, and an idealized discrimination variable $Y$, which for simplicity in our analysis replaces the usual log$_{10}$(S2/S1). The $S1$ distributions for the WIMP mass-dependent signal and  the various background sources are taken from the spectra shown in figure \ref{fi:finalbkg-s1}. 
Both ER and NR backgrounds are assumed to be Gaussian distributed in the discrimination variable $Y$: ERs have mean $=0$ and $\sigma =1$, while NRs have mean $=-2.58$ and $\sigma =0.92$. With this choice we reproduce the discrimination performance obtained in XENON100, namely:  
$99.5\%$ ER rejection at $50\%$ NR acceptance; 
 $99.75\%$ ER rejection at $40\%$ NR acceptance;
and $99.9\%$ ER rejection at $30\%$ NR acceptance \cite{xe100-instrument}.
In the profile likelihood analysis we use the whole data sample without applying any hard cut in the discrimination space $Y$, 
i.e. a marked Poisson distribution \cite{Conrad:2014nna}.


The expectation values for signal and backgrounds, considering a $2$\,year-long measurement with a $1$\,t FV, are summarized in table \ref{ta:xe1t}. 
The $S1$ range used in the analysis is ($3$,~$70$)~PE: 
the lower edge corresponds to the XENON100 S1 threshold, while the higher one marks the region where the ER background starts to be larger by more than an order of magnitude than the signal from a $100$~GeV/$c^2$ WIMP. On average, it corresponds to the NR energy range ($4$, $50$)~keV. 

\begin{table}[t!]
\centering
\normalsize
\begin{tabular}{|lcc|}
\hline 
\multicolumn{3}{|l|}{\textbf{Expectation values of events in XENON1T, in 2 t$\cdot$y exposure}} \\ \hline 
& No & 99.75\% ER \\ 
& discrimination & discrimination \\ \hline

\textbf{Signal} ($\mu_s$) & & \\
$6$ GeV/$c^2$ WIMP ($\sigma=2 \cdot 10^{-45}$ cm$^2$) & 0.68 & 0.27 \\
$10$ GeV/$c^2$ WIMP ($\sigma=2 \cdot 10^{-46}$ cm$^2$) & 4.65 & 1.86 \\
$100$ GeV/$c^2$ WIMP ($\sigma=2 \cdot 10^{-47}$ cm$^2$) & 7.13 & 2.85 \\
$1$ TeV/$c^2$ WIMP ($\sigma=2 \cdot 10^{-46}$ cm$^2$) & 8.85 & 3.54 \\
\textbf{Background} & & \\
Total ER ($\mu_{bER}$) & 1300 & 3.25 \\
\hspace{3mm}  NR from neutrons & 1.10 & 0.44 \\
\hspace{3mm}  NR from CNNS & 1.18 & 0.47 \\
Total NR ($\mu_{bNR}$) & 2.28 & 0.91 \\
\hline 
\end{tabular}
\caption{Number of expected events in XENON1T before and after $99.75\%$ ER discrimination ($40\%$ NR acceptance) in $1$~t fiducial volume and $2$~years of measurement. The $S1$ range is ($3$, $70$)~PE. } 
\label{ta:xe1t}
\end{table}

The main systematic uncertainty in the prediction of the signal and the NR background comes from the relative scintillation efficiency in LXe, $\mathcal{L}_\mathrm{eff}$. We adopt the $\mathcal{L}_\mathrm{eff}$ parameterization shown in figure 1 of \cite{xe100-run8}, using the median of several direct measurements as the central value and parameterizing the uncertainty by a Gaussian distribution. 
We extrapolated $\mathcal{L}_\mathrm{eff}$ also below 3 keV, where no direct measurements exist so far (although there are hints of non-vanishing $\mathcal{L}_\mathrm{eff}$ from the neutron calibration in LUX \cite{LUX-DD}):
the median value reaches zero at 1 keV and the 1$\sigma$ and 2$\sigma$ bands are increased to reflect the larger systematic uncertainty. 
We checked that the sensitivity is not significantly affected (at most 20\% at low WIMP masses, where the impact is the largest) if we adopt a uniform uncertainty parameterization between the $\pm 2\sigma$ bands below 3 keV, instead of the Gaussian one.
$\mathcal{L}_\mathrm{eff}$  and its uncertainty are parameterized with a single nuisance parameter $t_\mathcal{L}$, normally distributed with zero mean and unit variance. For each value of $t_\mathcal{L}$, we calculate the corresponding expected number of events and spectra for the NR background from neutrons and CNNS, and for the WIMP signal from each of the considered WIMP masses. We checked that the shape of the spectra is not significantly affected by the variation in $t_\mathcal{L}$, therefore only the variation in the expected number of events is considered in the analysis, as it was done also in \cite{xe100-pl}. The largest impact is observed for low energy signals, especially for CNNS and the low mass WIMPs, below $10$~GeV/$c^2$. Varying $\mathcal{L}_\mathrm{eff}$ to its $\pm 2 \sigma$ values produces a variation up to a factor 4 for CNNS and $6$~GeV/$c^2$ WIMPs, while for NR background from neutrons and $50$~GeV/$c^2$ WIMPs the variation is about $\pm 10\%$. 

We also included in the model other additional uncertainties, also treated as nuisance parameters in the likelihood. They include the charge yield $\mathcal{Q}_\mathrm{y}$, treated in the same way as $\mathcal{L}_\mathrm{eff}$ through the nuisance parameter $t_\mathcal{Q}$, and the systematic uncertainty on the prediction of the ER and NR backgrounds, assumed conservatively as $10\%$ and $20\%$, respectively, and parameterized with $t_{ER}$ and $t_{NR}$. We found them to be less relevant than the one coming from $\mathcal{L}_\mathrm{eff}$, at the level of a few percent in the final sensitivity with respect to the use of the $\mathcal{L}_\mathrm{eff}$  term only. 

We use an un-binned, extended likelihood defined as:
\begin{eqnarray}
  -2 ~ \ln ~L(\sigma; \, t_\mathcal{L}, t_\mathcal{Q}, t_{ER}, t_{NR}) &=& 2~ [~ \mu_s(\sigma; \, t_\mathcal{L}, t_\mathcal{Q}) + \mu_{b ER}(t_{ER}) + \mu_{b NR}(t_\mathcal{L}, t_\mathcal{Q}, t_{NR}) ~] \nonumber \\
   &-& ~ 2 ~ \sum_{i=1}^{n_{obs}} ~ \ln \{ ~ [ ~ \mu_s(\sigma; \, t_\mathcal{L}, t_\mathcal{Q}) \cdot f_s(S1_i) \cdot g_s(Y_i) ~ ]   \nonumber \\
   &+& ~ [ ~ \mu_{bER}(t_{ER}) \cdot f_{bER}(S1_i) \cdot g_{bER}(Y_i) ~ ]  \nonumber \\
   &+& ~ [ ~ \mu_{bNR}(t_\mathcal{L}, t_\mathcal{Q}, t_{NR}) \cdot f_{bNR}(S1_i) \cdot g_{bNR}(Y_i) ~ ] ~ \}  \nonumber \\
   &+& ~ [ ~ (t_\mathcal{L}-t_\mathcal{L}^0)^2 ~ + ~ (t_\mathcal{Q}-t_\mathcal{Q}^0)^2   \nonumber \\
   &+& ~ (t_{ER}-t_{ER}^0)^2 ~ + ~ (t_{NR}-t_{NR}^0)^2 ~]
\label{eq:LL}
\end{eqnarray}
where $\sigma$ is the spin-independent WIMP-nucleon cross section, $\mu_s$,  $\mu_{b ER}$ and  $\mu_{b NR}$ are the expectation values of the signal, the ER and NR background, respectively. $n_{obs}$ is the total number of observed events, $f(S1)$ and $g(Y)$ are the probability distributions in the two observables $S1$ and $Y$. The last term, $ \sum_{j}(t_{j}-t_{j}^0)^2$, with $j$ running over the four nuisance parameters, describes the Gaussian constraint on them, where $t_j^0$ is the expected value of the $j$-th nuisance parameter (e.g. $t_\mathcal{L}^0=0$ describes the median of $\mathcal{L}_\mathrm{eff}$). 
Note that both the WIMP signal and the NR background are affected by the uncertainty on $\mathcal{L}_\mathrm{eff}$ and $\mathcal{Q}_\mathrm{y}$.

Since we are interested in determining upper limits, we use the test statistic defined as: 

\begin{equation} \label{eq:test_stat_cases_PL}
q_\sigma =
\begin{cases}
-2 \, \text{ln} \frac{L(\sigma; \, \hat{\hat{t_j}})}{L(\hat{\sigma}; \, \hat{t_j})} & \text{if } \sigma \geq \hat{\sigma} \\
0                                                                     & \text{if } \sigma < \hat{\sigma} \\
\end{cases}
\end{equation}  
where $\hat{\sigma}$ and $\hat{t_j}$ are the maximum likelihood estimators (MLE), while $\hat{\hat{t_j}}$ is the conditional MLE obtained for the nuisance parameters at the fixed value of $\sigma$ under test.

In order to calculate the sensitivity, we perform Monte Carlo simulations ($10^4$ in each configuration) of the measurement process to obtain distributions of the test statistic under the background-only ($H_0$) and the signal hypotheses ($H_{\sigma}$, for all considered cross-sections). The $90\%$ C.L. upper limit is found using the CL$_s$ method, which prevents over-optimistic results due to under-fluctuations of the background. Following the nomenclature used in \cite{xe100-pl}, we require a $p'$-value equal to 0.1 for the $H_{\sigma}$ hypothesis to determine the upper limit $\sigma_{90}$. 
The sensitivity is then obtained by repeating these calculations for $10^4$ simulations of the background-only case, thereby obtaining the distribution of upper limits and in particular their median. It should be noted that we treat the nuisance parameters  as a random variable, i.e. for each MC experiment we draw also an estimate of $t_j^0$ according to the assumed Gaussian distribution. This corresponds to the unconditional ensemble discussed in \cite{ATLASnote}.

\begin{figure}[t!]
\centering\includegraphics[width=0.9\linewidth]{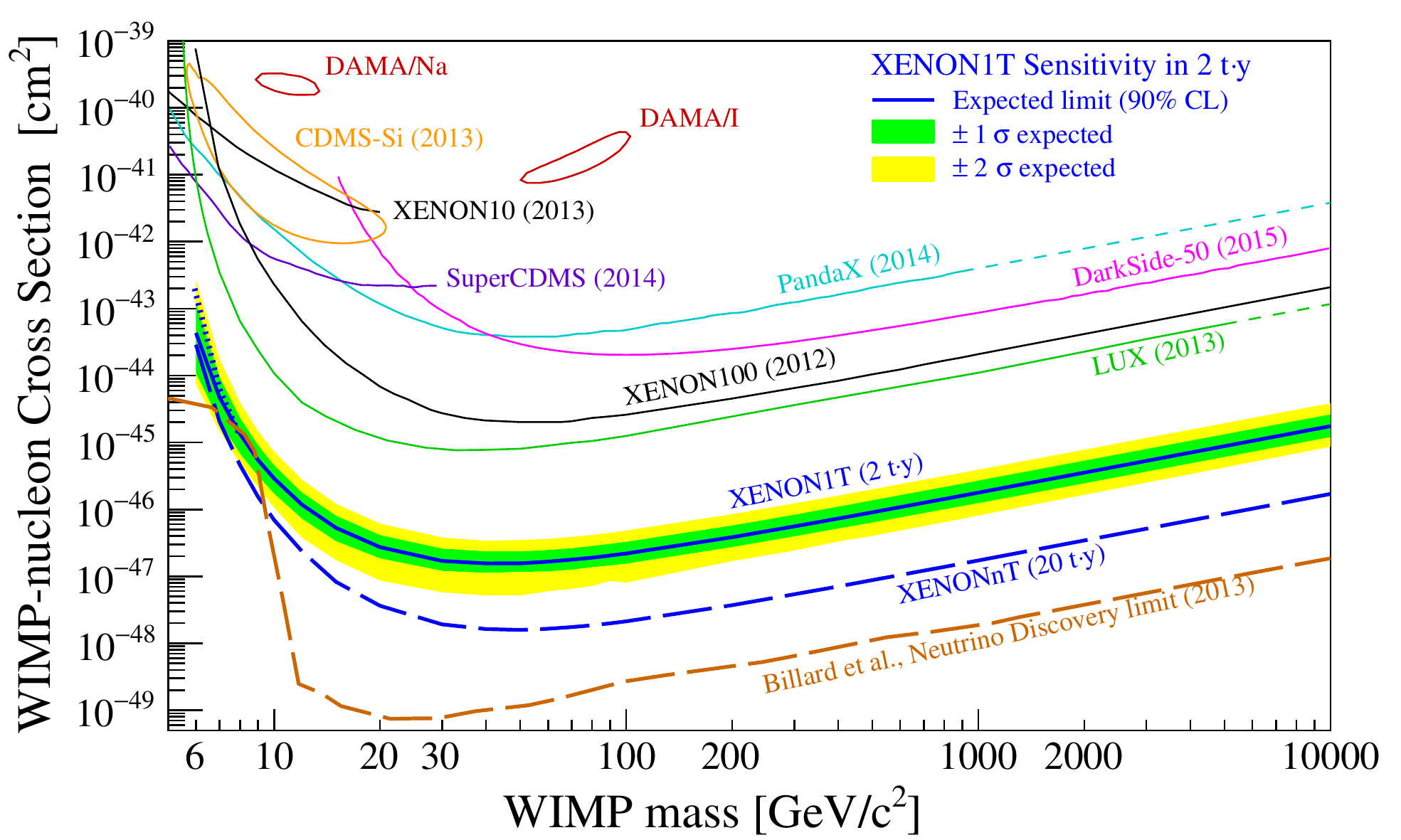}
\caption{XENON1T sensitivity (90\% C.L.) to spin-independent WIMP-nucleon interaction: the solid blue line represents the median value, while the $1\sigma$ and $2\sigma$ sensitivity bands are indicated in green and yellow respectively. The dotted blue line, visible at low WIMP masses, shows the XENON1T sensitivity assuming $\mathcal{L}_\mathrm{eff}=0$ below $3$~keV.  The XENONnT median sensitivity is shown with the dashed blue line.
The discovery contour of DAMA-LIBRA \cite{dama_savage2009} and CDMS-Si \cite{cdms-si2013} are shown, together with the exclusion limits of other experiments: XENON10 \cite{XENON10:2013}, SuperCDMS \cite{SuperCDMS2014}, 
PandaX \cite{PandaX2014}, DarkSide-50 \cite{DarkSide:2015}, XENON100 \cite{xe100-run10}, LUX \cite{LUX-results}. 
For comparison, with the dashed brown line we plot also the "neutrino discovery limit" from \cite{Billard}.}
\label{fi:xe1t-sens}
\end{figure}

\begin{figure}[t!]
\centering\includegraphics[width=0.9\linewidth]{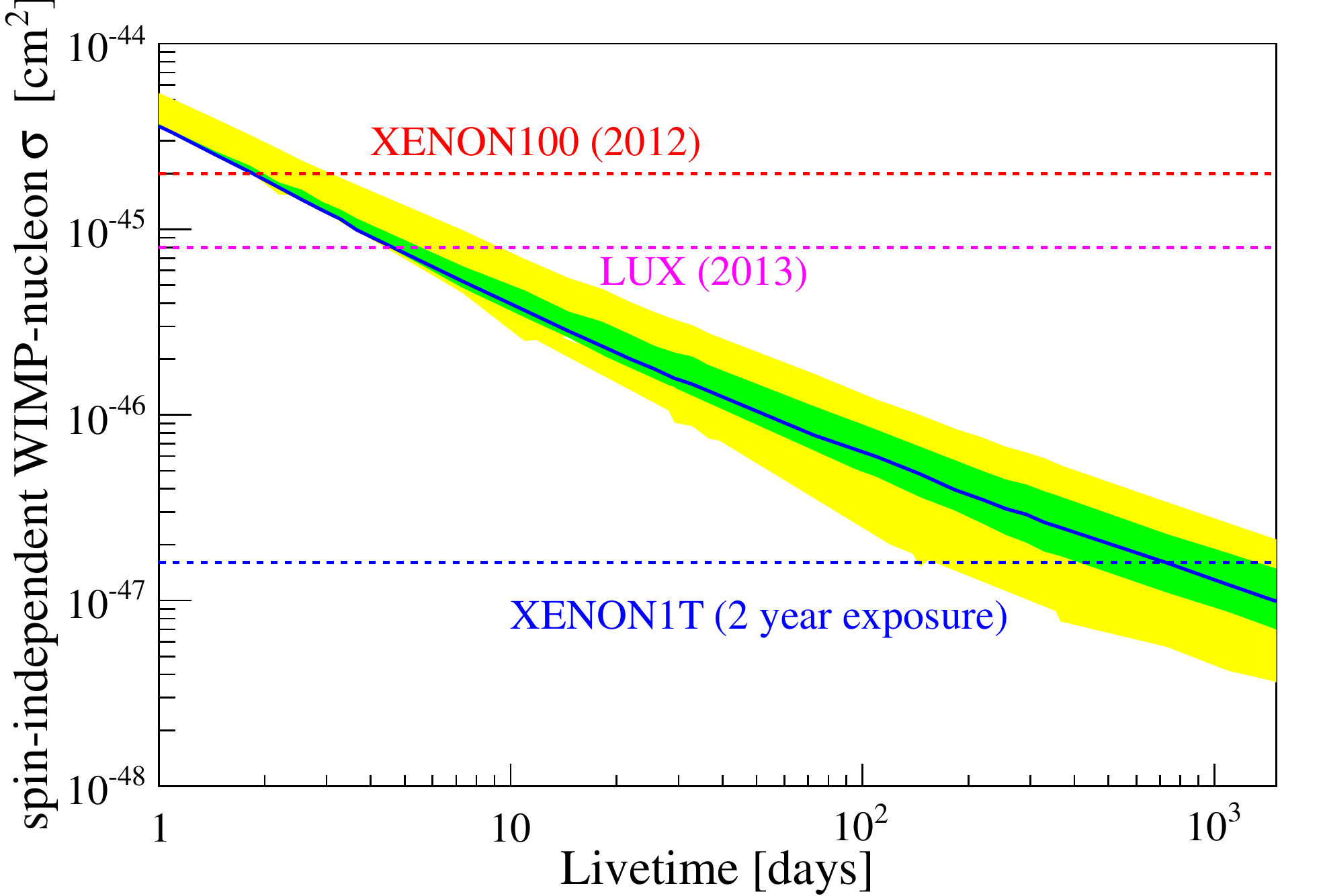}
\caption{XENON1T sensitivity (90\% C.L.) at $m_\chi=50$~GeV/$c^2$ in $1$~t FV as a function of the exposure time: the blue line represents the median value, and the $1\sigma$ and $2\sigma$ bands are indicated in green and yellow respectively. With the dashed red line we indicate the sensitivity of the XENON100 experiment \cite{xe100-run10}, with the dashed purple line that of LUX \cite{LUX-results}, and in blue the one obtained by XENON1T in  $2$~t$\cdot$y exposure.}
\label{fi:xe1t-sens-vs-time}
\end{figure}

Considering a XENON1T exposure of $2$ years in $1$~t FV, we obtain the result shown in figure \ref{fi:xe1t-sens}. The minimum is achieved at $m_\chi=50$ GeV/$c^2$ for a spin-independent WIMP-nucleon cross section of $1.6 \cdot 10^{-47}$ cm$^2$. The improvement with respect to the current best limit (from LUX \cite{LUX-results}) at the same WIMP mass is a factor $\sim$ 50. 
Assuming a cutoff in the emission model, with $\mathcal{L}_\mathrm{eff}$ set to zero below $3$~keV, the sensitivity decreases significantly only at low WIMP masses, below $10$ GeV/$c^2$, as it is shown with the blue dotted line. At $m_\chi=6$ GeV/$c^2$, the decrease in sensitivity with this pessimistic assumption is about a factor 3.
In figure \ref{fi:xe1t-sens-vs-time} we show the XENON1T sensitivity at $m_\chi=50$ GeV/$c^2$ as a function of time, assuming a 1 t FV. In about $2$ ($5$) days we can achieve the upper limit set by the XENON100 (LUX) experiment.

\subsection{Sensitivity projection towards XENONnT}
\label{S:XENONnT}

Many of the subsystems of the XENON1T experiment, e.g., water shield, Cherenkov muon veto, cryostat support, outer cryostat, LXe cooling, storage and purification systems, data acquisition system, etc., have been designed such that they can be used for an upgraded larger phase of the experiment, XENONnT, containing about $7$~tonnes of LXe. The TPC will be enlarged by ($20-30$)\% and the number of PMTs will increase to about $450$. Due to the improved self-shielding and capability to detect multiple scatters in the larger detector, it is more effective to define a FV in which the ER and NR backgrounds from the materials can be reduced to a negligible level. In addition, there are on-going R\&D studies to further improve the purification of the LXe target from the intrinsic contaminants $^{222}\rm{Rn}$ and $^{85}\rm{Kr}$.
Thus, we calculate the sensitivity of the XENONnT experiment assuming negligible ER and NR backgrounds from the detector materials, assuming $0.1$~$\mathrm{\mu}$Bq/kg of $^{222}\rm{Rn}$, and $0.02$~ppt of $^\mathrm{nat}\rm{Kr}$. With these assumptions, the main backgrounds come from ERs and NRs induced by solar neutrinos \footnote{
Another potential line of improvement is in the ER/NR discrimination, see e.g. \cite{Schumann:2015cpa}. However, in this study we conservatively assume the same discrimination considered for XENON1T. }.

The expected number of signal and background events for a total exposure of $20$~t$\cdot$y are summarized in table \ref{ta:xent}.
The sensitivity is presented in figure \ref{fi:xe1t-sens}: XENONnT will achieve a minimum spin-independent WIMP-nucleon cross section of $1.6 \cdot 10^{-48}$ cm$^2$ at $m_\chi$=$50$~GeV/$c^2$, with an improvement of an order of magnitude with respect to XENON1T. The results are very similar to those expected for the LZ experiment \cite{LZ-CDR:2015}.

\begin{table}[t!]
\centering
\normalsize
\begin{tabular}{|lcc|}
 \hline
\multicolumn{3}{|l|}{\textbf{Expectation values of events in XENONnT, in 20 t$\cdot$y exposure}} \\ \hline 
& No & 99.75\% ER \\ 
& discrimination & discrimination \\ \hline

\textbf{Signal} ($\mu_s$)  & & \\
$6$ GeV/$c^2$ WIMP ($\sigma=2 \cdot 10^{-46}$ cm$^2$) & 0.68 & 0.27 \\
10 GeV/$c^2$ WIMP ($\sigma=2 \cdot 10^{-47}$ cm$^2$) & 4.65 & 1.86 \\
100 GeV/$c^2$ WIMP ($\sigma=2 \cdot 10^{-48}$ cm$^2$) & 7.13 & 2.85 \\
1 TeV/$c^2$ WIMP ($\sigma=2 \cdot 10^{-47}$ cm$^2$) & 8.85 & 3.54 \\
\textbf{Background} & &\\
Total ER ($\mu_{bER}$) & 1000 & 2.5 \\
NR from neutrons & - & - \\
NR from CNNS ($\mu_{bNR}$) & 11.8 & 4.7 \\
\hline 
\end{tabular}
\caption{Number of expected events in XENONnT before and after $99.75\%$ ER discrimination ($40\%$ NR acceptance) in $20$~t$\cdot$y. The $S1$ range is ($3$, $70$)~PE. } 
\label{ta:xent}
\end{table}


\section{Summary and Conclusions}
\label{S:Summ}

We performed a detailed Monte Carlo simulation of the XENON1T experiment with a GEANT4 model. Considering the contaminations of the detector construction materials, measured through a screening campaign performed with Ge and mass spectrometry techniques, and the contaminants intrinsic to the LXe, 
we estimated both the ER and NR backgrounds. 

Selecting single scatter events in the ($1$, $12$)~keV range, assuming a $1$~t FV, the ER background rate is summarized in table \ref{ta:er-summary}. The most relevant contribution, about $85\%$ of the total ER background, comes from $^{222}\rm{Rn}$, while the one from the materials is of the same order of those coming from $^\mathrm{85}\rm{Kr}$ and elastic scattering of solar neutrinos ($\sim 5 \%$ each).
The total ER background is  $(1.80 \pm 0.15) \cdot 10^{-4}$ \dru, a factor $\sim 30$ lower than in XENON100 \cite{xe100-erbkg}, which corresponds to  $(720 \pm 60)$~\tyu $\,$ before applying any discrimination selection. 


The NR background has been studied in the energy region ($4$, $50$)~keV, which corresponds to the same S1 range used for ER when taking into account the different response of LXe to ER and NR.  Using the measured radio-activities of materials, we estimated a rate of $(0.6 \pm 0.1)$ \tyu $\,$ from radiogenic neutrons.  Due to the performance of the water Cherenkov muon veto, the background from muon-induced neutrons is reduced to less than $1 \cdot 10^{-2}$~\tyu. 
A different approach is needed for the NR background from coherent scattering of neutrinos: indeed their rate in the same energy region is very small,  $(1.8 \pm 0.3) \cdot 10^{-2}$~\tyu. Given their very steep energy spectrum, it is relevant to calculate their contribution after the conversion from energy into the signal seen in the detector, to correctly take into account the fluctuations due to the small number of detected photons at low energies.

The LCE for the S1 signal has been calculated with a MC simulation of the propagation of photons inside the TPC, considering the effects of the refractive index, the absorption length in LXe, the transparency of the various electrodes and the reflectivity of PTFE. Assuming realistic values for all these parameters, the resulting LCE averaged over the whole TPC active volume is $35\%$. This corresponds to a light yield at zero field of $7.7$~PE/keV ($4.6$~PE/keV at $530$~V/cm) at $122$~keV $\gamma$ energy. 

We have studied the WIMP signal and the expected backgrounds by converting energy depositions from ERs and NRs into observable signals, taking into account the detector resolution.
Considering as reference a $99.75\%$ ER discrimination with a corresponding $40\%$ NR acceptance, the background in the ($3$, $70$)~PE S1 range is $(1.62 \pm 0.15)$ \tyu $\,$ from ER, $(0.22 \pm 0.04)$ \tyu $\,$ from NR of radiogenic neutrons, and $(0.23 \pm 0.04)$ \tyu $\,$ from NR of neutrino coherent scattering. The uncertainties reflect only those coming from the knowledge of the sources of background, and not those from the LXe response which have been directly included in the sensitivity estimation.

We calculated the XENON1T sensitivity using the Profile Likelihood Ratio method, without any ER/NR discrimination cut. The main systematic uncertainty comes from $\mathcal{L}_\mathrm{eff}$, treated as a nuisance parameter affecting both the signal from WIMPs and the NR backgrounds. After a $2$~y measurement in $1$~t FV, we obtain the sensitivity shown in figure \ref{fi:xe1t-sens}, where the median value of the spin-independent WIMP-nucleon cross section reaches a minimum of $1.6 \cdot 10^{-47}$~cm$^2$ at m$_\chi$=50~GeV/$c^2$. 

We also estimated the sensitivity of XENONnT, a future upgrade to XENON1T, which will be hosted in the same experimental area and will contain up to $7$~t of LXe. Considering a FV where the ER and NR backgrounds from the detector materials are suppressed to a negligible level and assuming an improved purification from intrinsic contaminants, the most relevant background comes from ERs and NRs from solar neutrinos. Assuming a $20$~t$\cdot$y exposure, the sensitivity reaches $1.6 \cdot 10^{-48}$~cm$^2$ at m$_\chi$=50~GeV/$c^2$, an order of magnitude better than XENON1T.

In conclusion, with the XENON1T and XENONnT experiments we will be able to reach an unprecedented sensitivity to galactic dark matter particles, more than two orders of magnitude with respect to the currently running experiments. This will allow us to probe the region of electroweak parameter space favored by theoretical calculations in supersymmetric and other WIMP models \cite{BertoneBook, Strege:2012bt, Bagnaschi:2015eha}.




\appendix
\section{Appendix: XENON1T sensitivity with the {\it LUX2015} emission model}
\begin{figure}[t!]
\centering\includegraphics[width=0.49\linewidth]{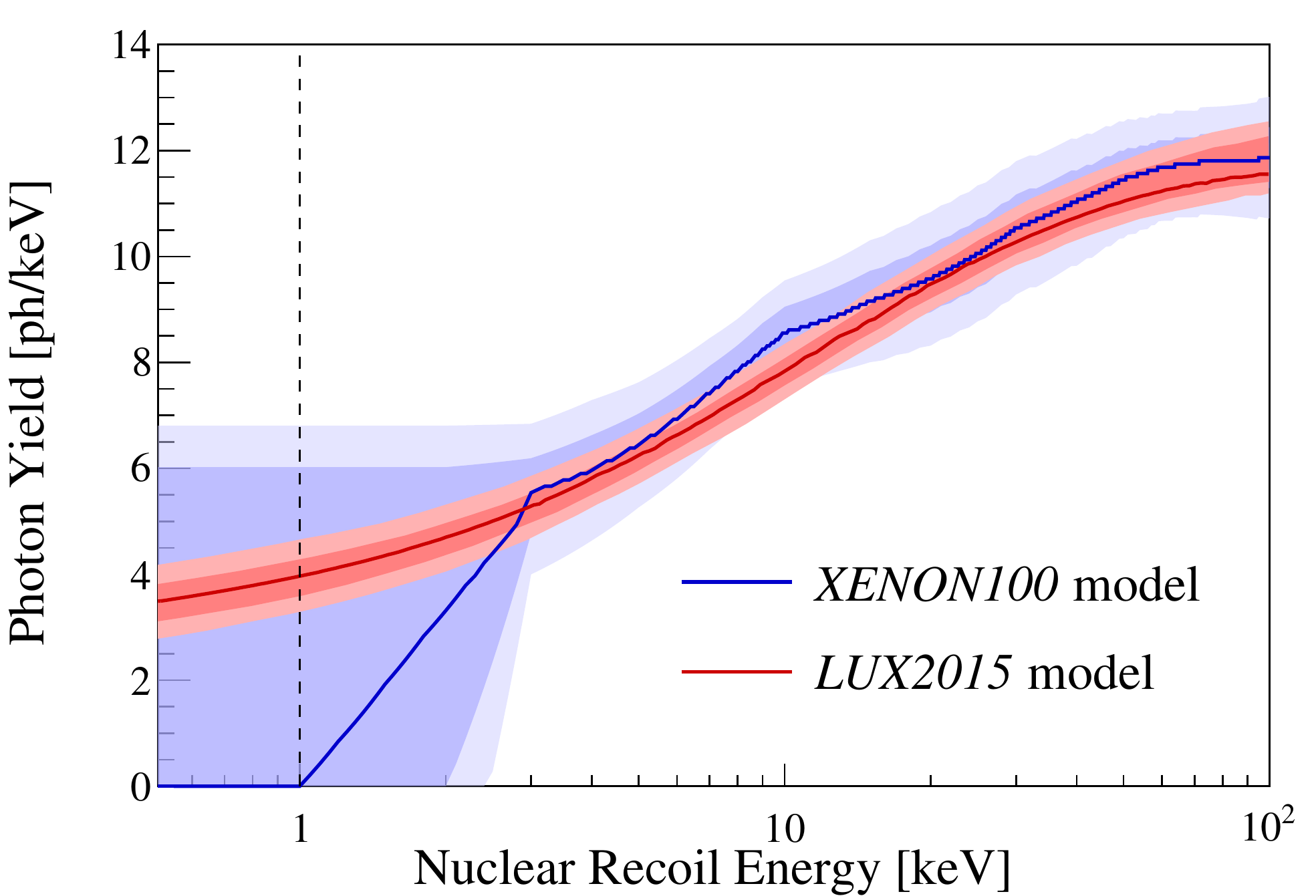}
\centering\includegraphics[width=0.49\linewidth]{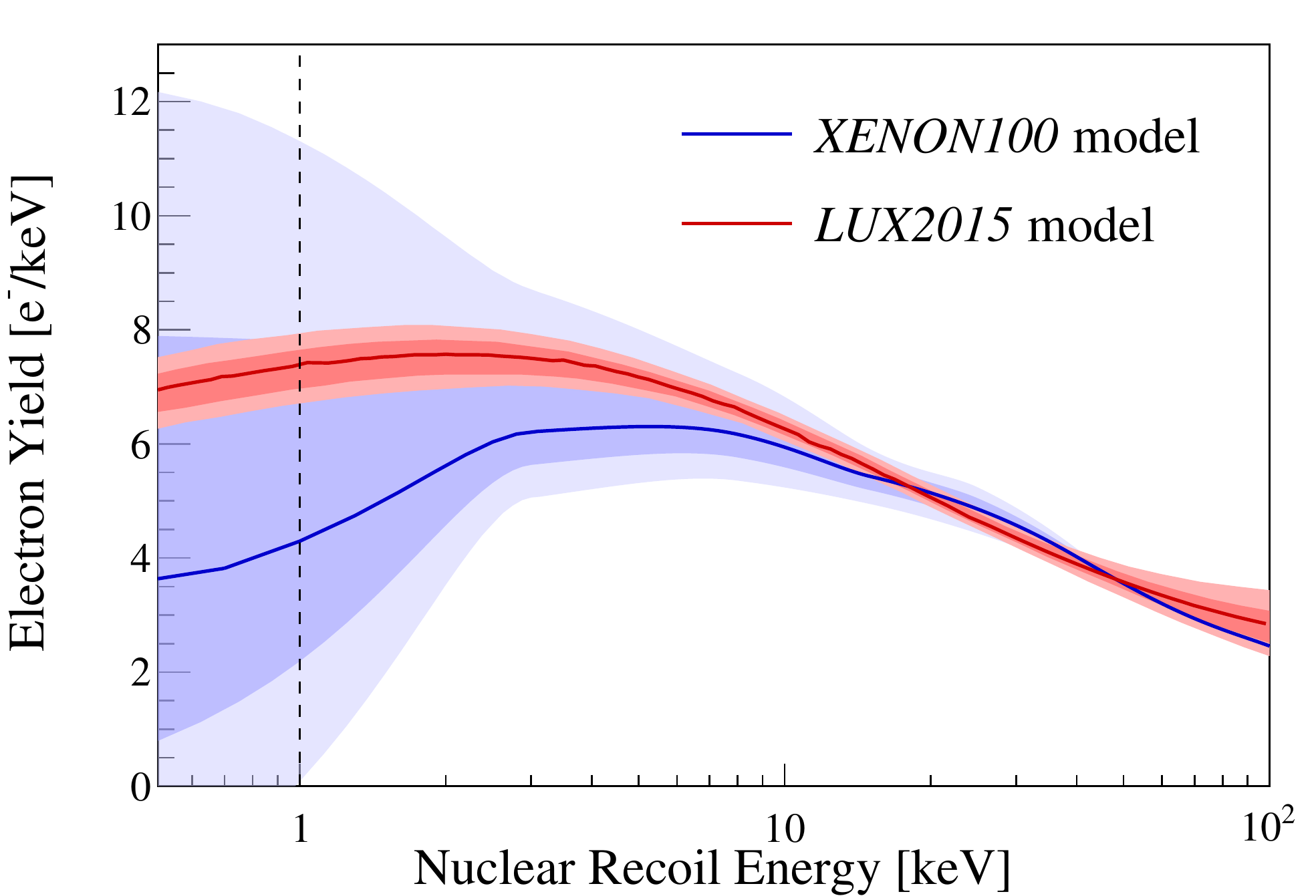}

\caption{Comparison of the LXe models for the emission of photons (left) and electrons (right) assumed in XENON100 (blue) and LUX (red). The thick line shows the median value, while the shaded areas represent the $1\sigma$ and $2\sigma$ bands. The vertical dashed line represents the cutoff energy below which the yields are assumed to be zero in both models.}
\label{fi:LUX2015-Ly}
\end{figure}

Recently, when we already were at the completion of this work, 
a new analysis of the data of the LUX experiment has been presented \cite{LUX:2015}. 
In that work a new model for the light and charge emission in LXe, driven by an {\it in situ} neutron calibration with a D-D source, was considered. In figure \ref{fi:LUX2015-Ly}, we compare the photon and electron yields measured in LUX (here called {\it LUX2015} model) to the one used in section \ref{S:Conv} (called {\it XENON100} model). For energies larger than $3$~keV, the photon yields are very similar, while at lower energies the  {\it LUX2015} model is significantly larger. Indeed, due to the lack of direct measurements below $3$~keV, in the {\it XENON100} model a photon yield  extrapolated down to zero at $1$~keV was conservatively assumed. The electron yield is larger in the  {\it LUX2015} model for most of the energy range of interest for WIMP-induced NRs.
With the new measurement, the systematic uncertainty in the yields is significantly decreased, as shown by the narrow $1\sigma$ and $2\sigma$ bands around the median value.
The higher yields allow the LUX experiment to obtain a large increase in sensitivity, in particular at low WIMP masses, with respect to their previous analysis \cite{LUX-results}.

The emission model in LXe impacts all LXe-based detectors in the same way, independent of detector-specific details. 
Therefore, a proper comparison of results or sensitivities must employ the same emission model in LXe.
Here we study the XENON1T sensitivity adopting the {\it LUX2015} model. 
Thus, we repeat the procedure described in section  \ref{S:Conv}, but with the new model and its uncertainties. As it was done in the LUX analysis, we also consider a cutoff at $1$~keV, below which the emission is set to zero.

\begin{table}[h!]
\centering
\normalsize
\begin{tabular}{|lcc|}
\hline 
\multicolumn{3}{|l|}{\textbf{Expectation values of events in XENON1T, in 2 t$\cdot$y exposure}} \\ \hline 
& {\it XENON100} & {\it LUX2015} \Tstrut \\ 
& model & model \\ \hline

\textbf{Signal} ($\mu_s$) & & \\
6 GeV/$c^2$ WIMP ($\sigma=2 \cdot 10^{-45}$ cm$^2$) & 0.68 & 2.72  \\
10 GeV/$c^2$ WIMP ($\sigma=2 \cdot 10^{-46}$ cm$^2$) & 4.65 & 5.96  \\
100 GeV/$c^2$ WIMP ($\sigma=2 \cdot 10^{-47}$ cm$^2$) & 7.13 & 7.13   \\
1 TeV/$c^2$ WIMP ($\sigma=2 \cdot 10^{-46}$ cm$^2$) & 8.85 & 8.85  \\
\textbf{Background} & & \\
Total ER ($\mu_{bER}$) & 1300 & 1300 \\
\hspace{3mm} NR from neutrons & 1.10 & 1.13  \\
\hspace{3mm} NR from CNNS & 1.18 & 5.36  \\
Total NR ($\mu_{bNR}$) & 2.28 & 6.49  \\
\hline 
\end{tabular}
\caption{Comparison of the number of expected events in XENON1T, considering the {\it XENON100} and  {\it LUX2015} emission models, before ER discrimination, in $1$~t fiducial volume and $2$~years of measurement. The $S1$ range is ($3$, $70$)~PE. } 
\label{ta:xe1t-lux2015}
\end{table}

\begin{figure}[t!]
\centering\includegraphics[width=0.9\linewidth]{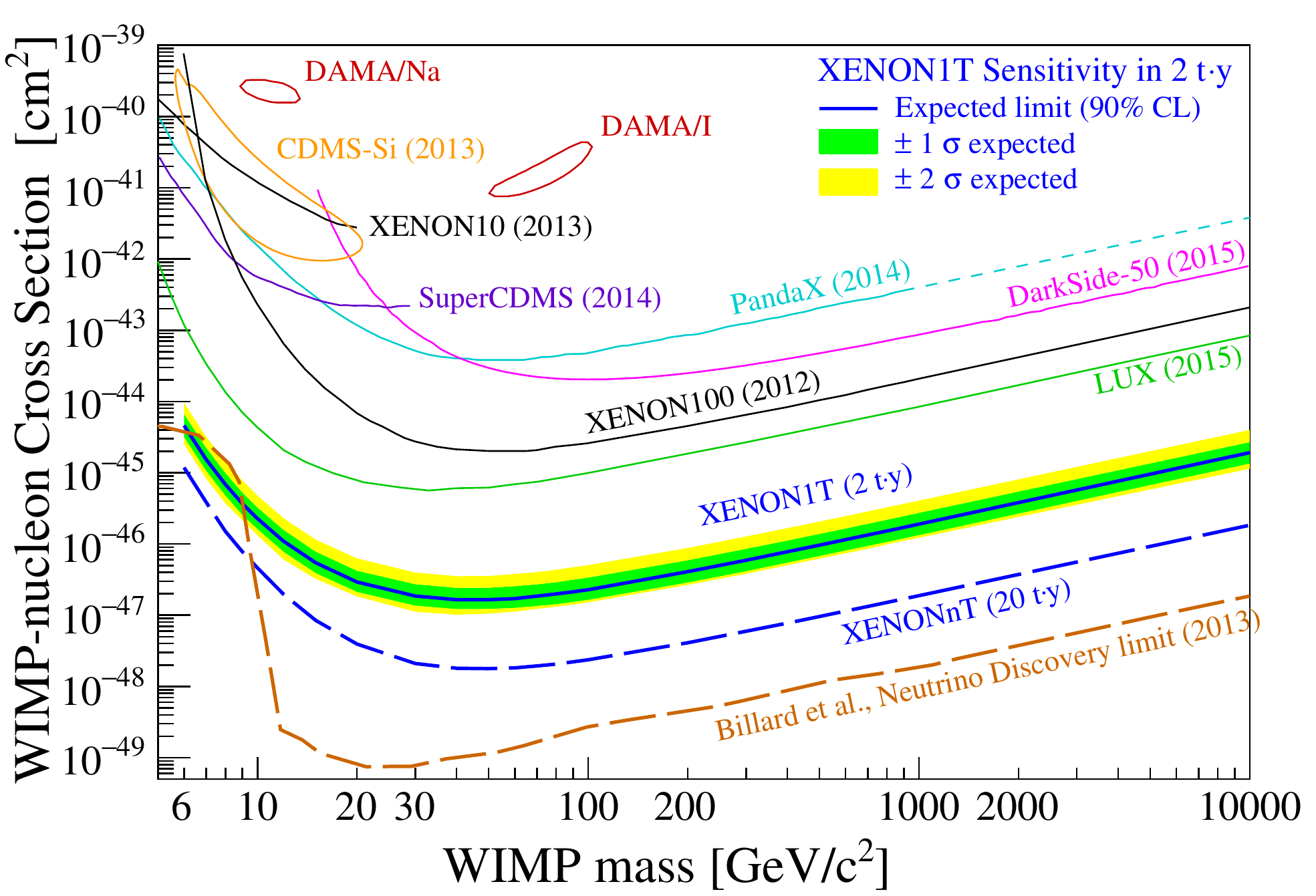}
\caption{XENON1T sensitivity (90\% C.L.) to spin-independent WIMP-nucleon interaction, calculated with the {\it LUX2015} emission model: the solid blue line represents the median value, while the $1\sigma$ and $2\sigma$ sensitivity bands are indicated in green and yellow respectively.  The XENONnT median sensitivity, also calculated with the {\it LUX2015} model, is shown with the dashed blue line.
The discovery contour of DAMA-LIBRA \cite{dama_savage2009} and CDMS-Si \cite{cdms-si2013} are shown, together with the exclusion limits of other experiments: XENON10 \cite{XENON10:2013}, SuperCDMS \cite{SuperCDMS2014}, 
PandaX \cite{PandaX2014}, DarkSide-50 \cite{DarkSide:2015}, XENON100 \cite{xe100-run10} and LUX with the 2015 re-analysis  \cite{LUX:2015}. 
For comparison, with the dashed brown line we plot also the "neutrino discovery limit" from \cite{Billard}.}
\label{fi:xe1t-sens-lux2015}
\end{figure}

The expected number of events in XENON1T for signal and backgrounds, in $2$~t$\cdot$y exposure, are summarized in table \ref{ta:xe1t-lux2015} and compared to those calculated with the {\it XENON100} model. We can see the increase in particular for the CNNS background ($\times 5$) and in the rates for low mass WIMPs ($\times 4$ at m$_\chi$=6~GeV/$c^2$).
The sensitivity of XENON1T, calculated assuming the {\it LUX2015} model and following the method described in section \ref{S:Sens}, is shown in figure \ref{fi:xe1t-sens-lux2015} and compared to the 2015 LUX results and to those of previous experiments. The minimum sensitivity is still at $1.6 \cdot 10^{-47}$~cm$^2$ at m$_\chi$=50~GeV/$c^2$, but the improvement at low mass WIMP is significant, about an order of magnitude at m$_\chi$=6~GeV/$c^2$ with respect to the one obtained with the {\it XENON100} model. In the same figure we also show the sensitivity of XENONnT, calculated in $20$~t$\cdot$y exposure with the assumptions described in section \ref{S:XENONnT}, here with the {\it LUX2015} model. 

\acknowledgments
We gratefully acknowledge support from:  
the National Science Foundation, 
Swiss National Science Foundation, 
Bundesministerium f\"ur Bildung und Forschung, 
Max Planck Gesellschaft, 
Foundation for Fundamental Research on Matter, 
Weizmann Institute of Science, I-CORE,
Initial Training Network Invisibles (Marie Curie Actions, PITN- GA-2011-289442),
Fundacao para a Ciencia e a Tecnologia, 
Region des Pays de la Loire, 
Knut and Alice Wallenberg Foundation, 
and Istituto Nazionale di Fisica Nucleare. 

We are grateful to Laboratori Nazionali del Gran Sasso for hosting and supporting the XENON project.



   \bibliographystyle{JHEP}
        \bibliography{mybibliography}

\end{document}